\documentclass[11pt]{article}
\usepackage{amssymb}
\usepackage{subfig}
\usepackage{color}
\usepackage{epsfig,amssymb,amsfonts,amsmath,graphicx,dsfont,cite,xfrac}
\usepackage{authblk}
\definecolor{mygray}{gray}{0.5}
\usepackage{cite}
\usepackage[colorlinks=true,linkcolor=blue,citecolor=red]{hyperref}
\usepackage{multirow}

\usepackage{tikz,pgfplots}
\usetikzlibrary{decorations.pathreplacing}
\usetikzlibrary{shapes.multipart}
\usetikzlibrary{positioning}
\usetikzlibrary{matrix}
\usetikzlibrary{shapes,calc,arrows}
\providecommand{\keywords}[1]
{
  \small	
  \textbf{\textit{Keywords---}} #1
}

\newcommand{\be}{\begin{equation}}
\newcommand{\ee}{\end{equation}}
\newcommand{\bea}{\begin{eqnarray}}
\newcommand{\eea}{\end{eqnarray}}

\title{Landau levels and snake states of pseudo-spin-1 Dirac-like electrons in gapped Lieb lattices}

\author[${}$]{V. Jakubsk\'y and K. Zelaya}
\affil[${}$]{\footnotesize Nuclear Physics Institute, Czech Academy of Science, 250 68 \v{R}e\v{z}, Czech Republic}

\date{}

\begin{document}

\maketitle

\begin{abstract}
This work reports the three-band structure associated with a Lieb lattice with arbitrary nearest and next-nearest neighbors hopping interactions. For specific configurations, the system admits a flat band located between two dispersion bands. Three inequivalent Dirac valleys are identified so that the quasi-particles are effectively described by the spin-1 Dirac-type equation. Under external homogeneous magnetic fields, the Landau levels are exactly determined as the third-order polynomial equation for the energy can be solved using Cardano's formula. It is also shown that an external anti-symmetric field promotes the existence of current-carrying states, so-called snake states, confined at the interface where the external field changes its sign.
\end{abstract}


%%%%%%%%%%%%%%%%%%%%%%%%%%%%%%%%%%%%%%%%%%%%%%%
\section{Introduction}
The rise of Dirac materials has paved the way to further understand and study the dynamics of relativistic (high-energy) fermions in low-energy excitations in condensed matter physics~\cite{Weh14}. In this regard, graphene, a honeycomb-shaped atomic-carbon lattice, is perhaps the most well-known system in condensed matter physics where lowe-energy electrons have properties akin to that of relativistic electrons described by the Dirac equation. The energy relations form dispersion bands that approximate to the so-called \textit{Dirac cones}. This has allowed testing relativistic properties of fermions such as Klein paradox in relatively simple condensed matter experiments~\cite{Kat06}. For a detailed overview of recent experimental and theoretical advances in monolayer and bilayer graphene, see~\cite{Aok14}.

There are indeed further two-dimensional lattice geometries associated with or beyond the honeycomb-shaped one that exhibits properties of relativistic Dirac electrons. The latter includes the well-known $\alpha-T3$ lattice~\cite{Dey19} that contains graphene and \textit{dice} lattices as special cases~\cite{Iur19}. The kagome lattice is another example of a honeycomb-shaped structure, even though its atoms are arranged differently from graphene~\cite{Mek03}. In contradistinction to the previous systems, the Lieb lattice~\cite{Apa10,Gol11,Slo17} constitutes a two-dimensional system not associated with a honeycomb geometry; instead, it has atoms periodically placed in the vertices of a square array, as well as in the center of each edge of such squares. Besides their geometrical structure, they are characterized by the number of atoms per unit cell. This key feature defines the pseudo-spin of the effective Dirac equation. The previous systems are examples of pseudo-spin-1 Dirac materials.

In particular, tight-binding models of Lieb lattices with nearest-neighbor interactions have been discussed in the context of optical lattices~\cite{Shen10,Guz14,Vic14,Die16}, as well as in electronic lattices formed by surface state electrons in Cu(111), as recently reported by Slot \textit{et al.} in~\cite{Slo17}. For next-nearest neighbor interactions, there are reports in magnon lattices under external periodically-driven fields~\cite{Owe18}. A further generalization of a system with flat bands has been studied in~\cite{Lim20} as a one-parameter model that transits from Lieb to Kagome lattices. In these cases, the system exhibits a three-band structure, whose solutions around the Dirac valleys are solved by using the \textit{partitioning method}~\cite{Low51}, which reduces the $3\times 3$ Dirac equation into an effective $2\times 2$ one for suitable energies. Other systems with three-bands have been considered in~\cite{Gol11} with next-neighbor interaction under the influence of external homogeneous magnetic fields, where the resulting energy equation has been solved approximately for the small spin-orbit interactions.

This work considers a Lieb lattice with nearest and next-nearest neighbor interactions with arbitrary hopping amplitude. To make our model more general, we allow the next-nearest interaction to hold a complex phase, which is achieved in analogy to the dimerized interactions of Haldane~\cite{Hal88} in graphene. Such a phase enriches the band structure of the lattice in the free-field configuration, as it changes the location of the Dirac valley and permits the emergence of flat bands either in between two dispersion bands or above (below) them. Furthermore, the hopping amplitude of the next-nearest neighbor allows the opening of a gap in the dispersion relations. With the aid of the \textit{Peierls transformation}~\cite{Pei29,Blo28}, we introduce and study the effects of external magnetic fields on the Lieb lattice. Although such a transformation was originally proved to hold for metallic single-atom cells under the influence of external electric potentials, it was shown that it also applies to non-metal multi-atom cell lattices under the action of external magnetic fields~\cite{Sla49,Lut51}. The latter allows accounting for external vector potentials minimally coupled to the Dirac Hamiltonian when expanded around the proper Dirac valleys. 

For homogeneous external magnetic fields, localized electrons and their related Landau levels are usually determined by either reducing the $3\times 3$ effective Dirac equation by one order~\cite{Lim20} when expanded for low-energy configurations\footnote{This is achieved using the \textit{partition method}~\cite{Low51}.}, or by considering small enough spin-orbit interactions~\cite{Gol11}. Here, we show that exact Landau levels can be determined with great generality through Cardano's formula~\cite{Olv10} without resorting to any approximation. Additionally, we consider an external anti-symmetric magnetic field where the new phenomena emerge at the interface where the field changes its sign. Unlike the homogeneous case, there exist current-carrying states with non-null group velocity in the direction parallel to the interface. This class of confinement was firstly discussed by M\"uller for two-dimensional electron gases~\cite{Mue92}, and has been found and studied in graphene in the form of \textit{snake states}~\cite{Oro08,Gho08,Liu15}.

The manuscript is structured as follows. The tight-binding model and the associated free-particle band structure are discussed in Sec.~\ref{sec:Lieb}, where the multiple allowed Dirac valleys are identified and classified for the spin-orbit coupling case. The influence of homogeneous external fields and the related Landau levels are calculated exactly through the Cardano formula for every Dirac valley in Sec.~\ref{sec:spin1}. In Sec.~\ref{sec:snake}, an anti-symmetric external magnetic field is considered to allow the confinement of electrons in the form of snake states parallel to the magnetic field discontinuity. A further discussion about other values of the phase hopping is presented in Sec.~\ref{sec:lambda}.  
%------------------------------>

\section{Dirac points of the three-band Lieb lattice}
\label{sec:Lieb}
\begin{figure}
\centering
\subfloat[][]{\includegraphics[width=0.32\textwidth]{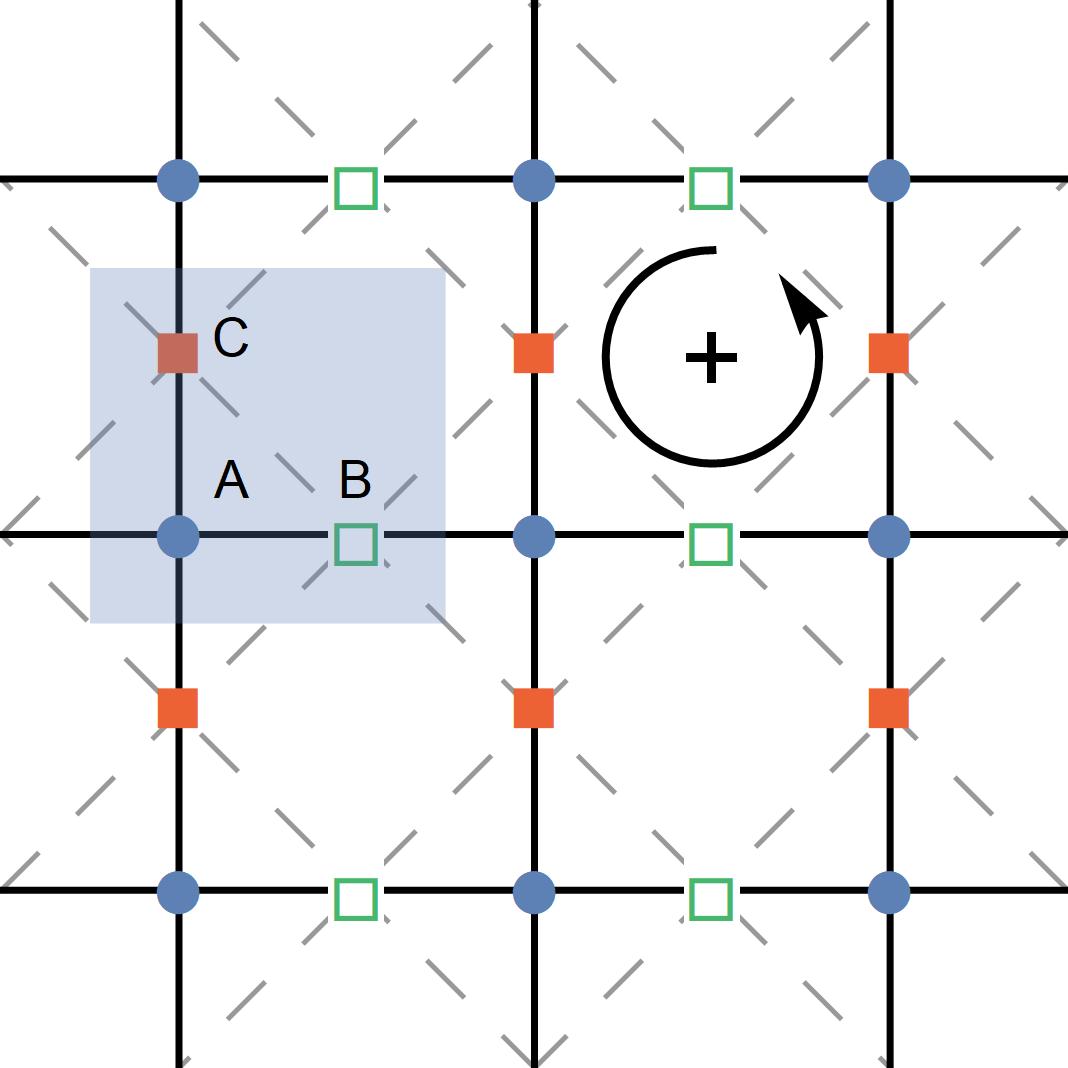}
\label{fig:Lieb-1}}
\hspace{2mm}
\subfloat[][]{\includegraphics[width=0.3\textwidth]{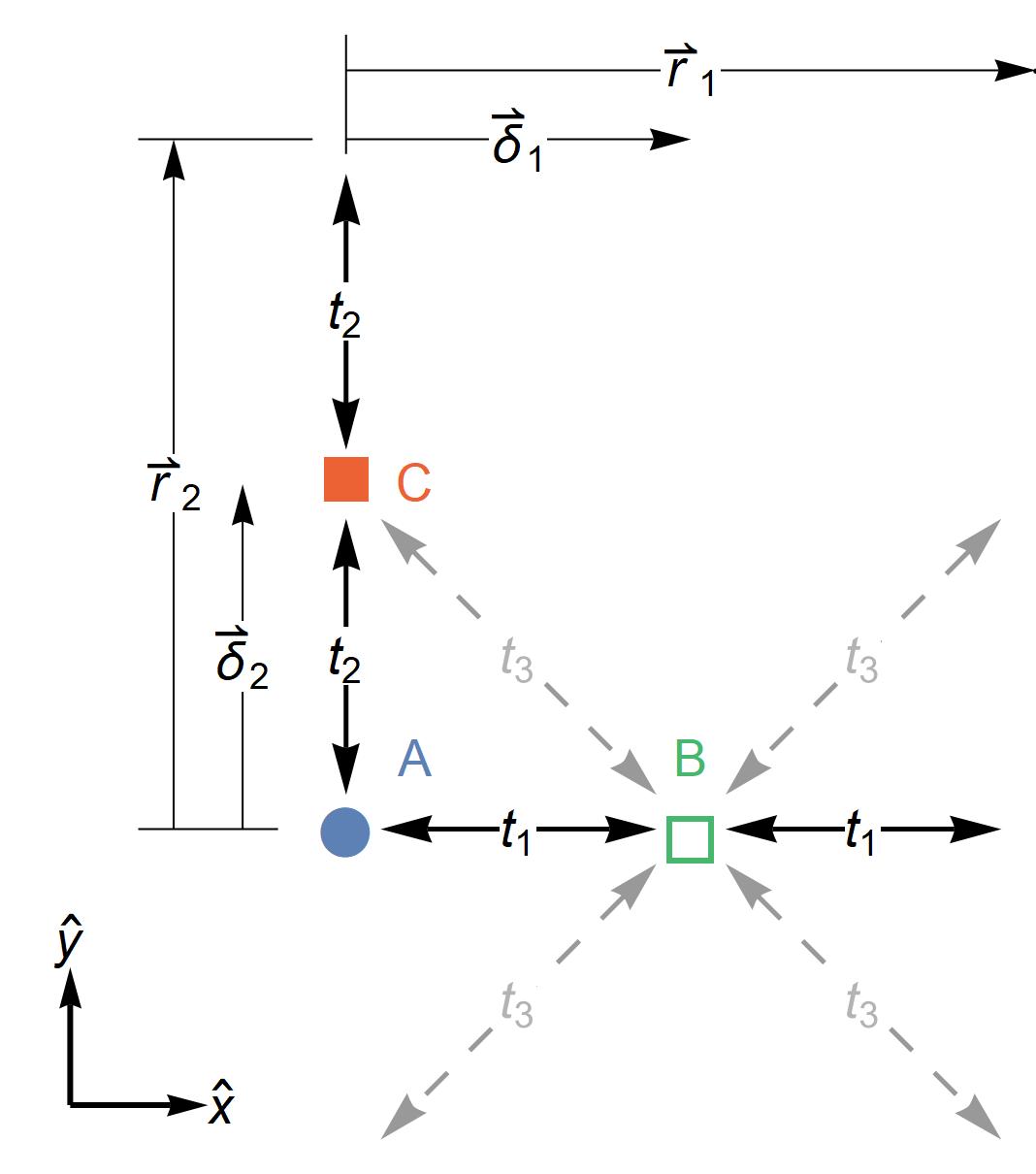}
\label{fig:Lieb-2}}
\hspace{2mm}
\subfloat[][]{\includegraphics[width=0.3\textwidth]{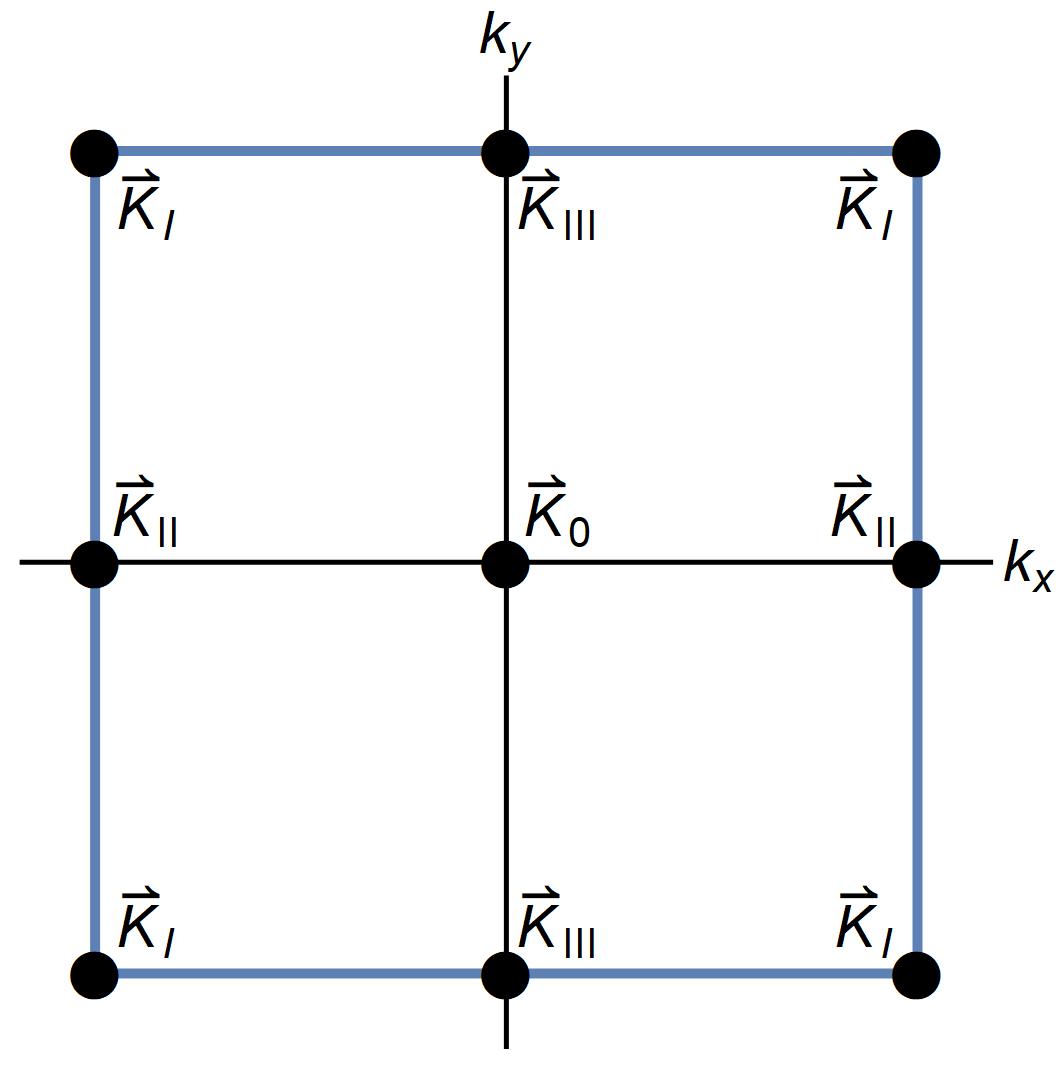}
\label{fig:Lieb-3}}
\caption{(a) Lieb lattice composed by the atoms $A$ (blue-filled-circle), $B$ (green-square), and $C$ (red-filled-square). The open circle indicates the direction of the phase hopping parameter for the next-nearest neighbor (NNN) $B-C$. The shadowed area depicts a unitary cell. (b) Composition of the shadowed unit cell in (a), where the unit displacement vectors $\vec{\delta}_{1}=a\hat{x}$ and $\hat{\delta}_{2}=a\hat{y}$ connect the atom $A$ with $B$ and $C$, respectively. The corresponding nearest neighbor (NN) hopping parameters are denoted by $t_{1}$, $t_{2}$, and $t_{3}$. (c) First Brillouin zone in the $\vec{k}$-space and the corresponding Dirac points $\{\vec{K}_{\operatorname{I}}=\left(\frac{\pi}{2a},\frac{\pi}{2a}\right),\vec{K}_{\operatorname{II}}=\left(\frac{\pi}{2a},0\right),\vec{K}_{\operatorname{III}}=\left(0,\frac{\pi}{2a}\right)\}$. The equivalent points are generated through the invariant translation vector $\vec{K}=(\frac{n\pi}{a},\frac{m\pi}{a})$ with $n,m\in\mathbb{Z}$. $\vec{K}_{0}=(0,0)$ denotes the origin in the $\vec{k}$-space.}
\label{fig:Lieb}
\end{figure}

The Lieb lattice is characterized by a two-dimensional rectangular array. The atoms are placed at the corners of each square and at the midpoint on each side of the square. The structure of the lattice is depicted in Fig.~\ref{fig:Lieb-1}. The separation between two nearest atoms is $a$, whereas the length of each side of the square is $\ell=2a$. Therefore, the primitive cell of the crystal has three atoms and it can reconstruct the whole lattice with the use of the primitive translation vectors $\vec{r}_{1}=2a\hat{x}$ and $\vec{r}_{2}=2a\hat{y}$. We introduce the lattice vector $\vec{\delta}_{1}=a \hat{x}=\vec{r}_{1}/2$ and $\vec{\delta}_{2}=a \hat{y}=\vec{r}_{2}/2$ that connect the atoms on site $A$ to those on the sites $B$ and $C$, respectively (see Fig.~\ref{fig:Lieb-2}). In this form, taking any atom $A$ as a reference, we can locate any other atoms $A$, $B$, and $C$ on the lattice through the translation vectors $\mathbf{R}_{A}=n_{1}\vec{r}_{1}+n_{2}\vec{r}_{2}$, $\mathbf{R}_{B}=\mathbf{\vec{R}}_{A}+\vec{\delta}_{1}$, and $\mathbf{R}_{C}=\vec{R}_{A}+\vec{\delta}_{2}$, respectively, with $n_{1},n_{2}\in\mathbb{Z}$. 

The reciprocal space is spanned by the translation vectors of the reciprocal space $\hat{r}_{k_{1}}$ and $\hat{r}_{k_{2}}$ constructed such that $\hat{r}_{p}\cdot\hat{r}_{k_{q}}=2\pi\delta_{p,q}$. %, defined through the reciprocal vectors $\vec{k}=k_{x}\hat{x}+k_{y}\hat{y}$, is formed by a square lattice with side lengths $\frac{\pi}{a}$. The latter follows from translation vectors in the reciprocal space
This leads to $\hat{r}_{k_{1}}=\frac{\pi}{a}\hat{x}$ and $\hat{r}_{k_{2}}=\frac{\pi}{a}\hat{y}$. The first Brillouin zone, constructed from the Wigner-Seitz rule, restricts to the region composed by $k_{x}\in[-\frac{\pi}{2a},\frac{\pi}{2a}]$ and $k_{y}\in[-\frac{\pi}{2a},\frac{\pi}{2a}]$. See Fig.~\ref{fig:Lieb-3}.

%This lattice has associated a square primitive cell with three atoms per cell as depicted in Fig.~\ref{fig:Lieb-1-b}. Henceforth 
We suppose that there is a single electron in each site that contributes to the electronic properties of the crystal. The corresponding wave function on site $\mathbf{R}_X$, $X=A,B,C$, can be written with the use of the corresponding creation operator as $|\mathbf{R}_X\rangle=\mathcal{C}_{\mathbf{R}_X}^\dagger|0\rangle$.  The probability that the electrons will skip from one site to the other one due to the inter-atomic interactions is encoded in the hopping parameters $t_{1}$ and $t_{2}$ that correspond to the nearest neighbor (NN) $A-B$ and $A-C$ transitions, respectively. The next-nearest neighbor (NNN) transition $B-C$ is reflected by $t_{3}e^{i\lambda}$. Here, the phase hopping $\lambda$ is taken positive if the hopping occurs in counter-clockwise direction, see Fig.~\ref{fig:Lieb}. The phase hopping $\lambda$ emerges from time-independent periodic gauge fields in honeycomb lattices~\cite{Hal88}, or through the coupling between spin and $s,p$ orbitals of different neighbors (spin-orbit coupling)~\cite{Kon10}. This also appears as the result of external time-dependent driven fields in photonic Lieb lattices~\cite{Lon17}, and magnon Lieb and Kagome lattices~\cite{Owe18}.

The tight-binding Hamiltonian of the Lieb lattice that describes both NN and NNN interactions writes as
%\begin{equation}
%H=-\sum_{\langle n,m\rangle}\left(t_{1}\mathcal{C}^{\dagger}_{A;n,m}\mathcal{C}_{B;n,m}+t_{2}\mathcal{C}^{\dagger}_{A;n,m}\mathcal{C}_{C;n,m}+ \right) - \sum_{\langle\langle n,m\rangle\rangle}t_{3}e^{i\mu \lambda}\mathcal{C}^{\dagger}_{B;n,m}\mathcal{C}_{C;m,n} +h.c. \, ,
%\label{TB-1}
%\end{equation}
\begin{equation}
H=-\sum_{\mathbf{R}_{A}}\sum_{j=1}^2\left(t_{j}\mathcal{C}^{\dagger}_{\mathbf{R}_A+\delta_j}\mathcal{C}_{\mathbf{R}_A}+t_{j}\mathcal{C}^{\dagger}_{\mathbf{R}_A-\delta_j}\mathcal{C}_{\mathbf{R}_A} \right) - \sum_{\epsilon_1,\epsilon_2=\pm 1}\sum_{\mathbf{R}_C}t_{3}e^{i\mu \lambda}\mathcal{C}^{\dagger}_{\mathbf{R}_C+\epsilon_1\delta_1+\epsilon_2\delta_2}\mathcal{C}_{\mathbf{R}_C} +h.c. \, ,
\label{TB-1}
\end{equation}
where the first term (together with its hermitian conjugate one) corresponds to NN and the second term (together with its h.c.) to NNN interaction. Also, $\mu=+1$ if the hopping between next-nearest neighbors occurs in counter-clockwise direction, and $\mu=-1$ otherwise\footnote{We have used the phase hopping direction in the opposite direction to that of Haldane~\cite{Hal88}.}. Remark that tight-binding model of Lieb lattices with only nearest neighbor interactions ($t_3=0$) has been discussed in the context of optical~\cite{Apa10,Shen10,Guz14,Vic14,Die16} and electronic~\cite{Slo17} lattices, whereas next-nearest neighbor intereactions ($t_{3}\neq 0$) in magnon lattices in~\cite{Owe18}.

It is customary to write the Hamiltonian~\eqref{TB-1} in the basis of the three Bloch wave functions\footnote{There holds $T_{r_{j}}|X\rangle=\sum_Xe^{-ik\mathbf{R}_X}T_{r_j}|\mathbf{R}_X\rangle=\sum_Xe^{-ik\mathbf{R}_X}T_{r_j}|\mathbf{R}_X+r_j\rangle=e^{i k r_j}|X\rangle$, $X=A,B,C$, where $T_{r_j}$ is the operator of translation by the vector $r_j$. } $|X\rangle=\sum_Xe^{-i\vec{k}\cdot\mathbf{R}_X}|\mathbf{R}_X\rangle$ of fixed quasi-momentum $\vec{k}=k_{x}\hat{x}+k_{y}\hat{y}$ that takes values from the first Brillouin zone, which is fixed by the vectors $\vec{T}_{k_x}=\frac{\pi}{a}\hat{x}$ and $\vec{T}_{k_{y}}=\frac{\pi}{a}\hat{y}$. It takes form of the $3\times 3$ constant matrix acting on the three-component spinors $\boldsymbol{\Psi}(\vec{k})=(\psi_{A},\psi_{B},\psi_{C})^{T}$. It reads explicitly as
\begin{equation}
\mathcal{H}(\vec{k})=
\begin{pmatrix}
0 & -2t_{1}\cos\left(ak_{x}\right) & -2t_{2}\cos\left(ak_{y}\right) \\
-2t_{1}\cos\left(ak_{x}\right) & 0 & -2t_{3}f^{*}(\vec{k},\lambda) \\
-2t_{2}\cos\left(ak_{y}\right) & -2t_{3}f(\vec{k},
\lambda) & 0
\end{pmatrix}
\label{TB-2}
\end{equation}
with $f(\vec{k},\lambda)=2\cos\left(ak_{x}\right)\cos\left(ak_{y}\right)\cos\lambda-2i\sin\left(ak_{x}\right)\sin\left(ak_{y}\right)\sin\lambda$. 
Its eigenvalues are defined as the roots of the characteristic equation
\begin{multline}
w^{3}(\vec{k})-4w(\vec{k})\left(t_{1}^{2}\cos^2\left(ak_{x}\right)+t_{2}^{2}\cos^2\left(ak_{y}\right)+t_{3}^{2}\vert f(\vec{k},\lambda)\vert^{2}\right)+\\
16t_{1}t_{2}t_{3}\cos(ak_{x})\cos(ak_{y})\operatorname{Re}[f(\vec{k},\lambda)]=0.
\label{characteristic}
\end{multline}
It can be solved analytically via Cardano's formula, giving rise to three real functions $w_{j}=w_{j}(k)$, $j=1,2,3$ that form dispersion relations,
\begin{equation}\label{genericE}
w_j(\vec{k})=\left(-\frac{q}{2}+\sqrt{\frac{q^2}{4}+\frac{p^3}{27}}\right)^{\frac{1}{3}}+\left(-\frac{q}{2}-\sqrt{\frac{q^2}{4}+\frac{p^3}{27}}\right)^{\frac{1}{3}}
\end{equation}
where the cubic roots have to be selected such that 
$\left(-\frac{q}{2}+\sqrt{\frac{q^2}{4}+\frac{p^3}{27}}\right)^{\frac{1}{3}}\left(-\frac{q}{2}-\sqrt{\frac{q^2}{4}+\frac{p^3}{27}}\right)^{\frac{1}{3}}=-\frac{p}{3}$, see appendix A for more details.
The quantities $p$ and $q$ have the following explicit form  
\begin{align}
p&=-4\left(t_{1}^{2}\cos^2\left(ak_{x}\right)+t_{2}^{2}\cos^2\left(ak_{y}\right)+t_{3}^{2}\vert f(\vec{k},\lambda)\vert^{2}\right)\\
q&=16t_{1}t_{2}t_{3}\cos(ak_{x})\cos(ak_{y})\operatorname{Re}[f(\vec{k},\lambda)].
\end{align}

\subsection{Spin-orbit coupling ($\lambda=\frac{\pi}{2}$)}
\label{subsec:spin-orbit-1}

\begin{figure}
\centering
\subfloat[][$t_{3}=0$]{\includegraphics[width=0.24\textwidth]{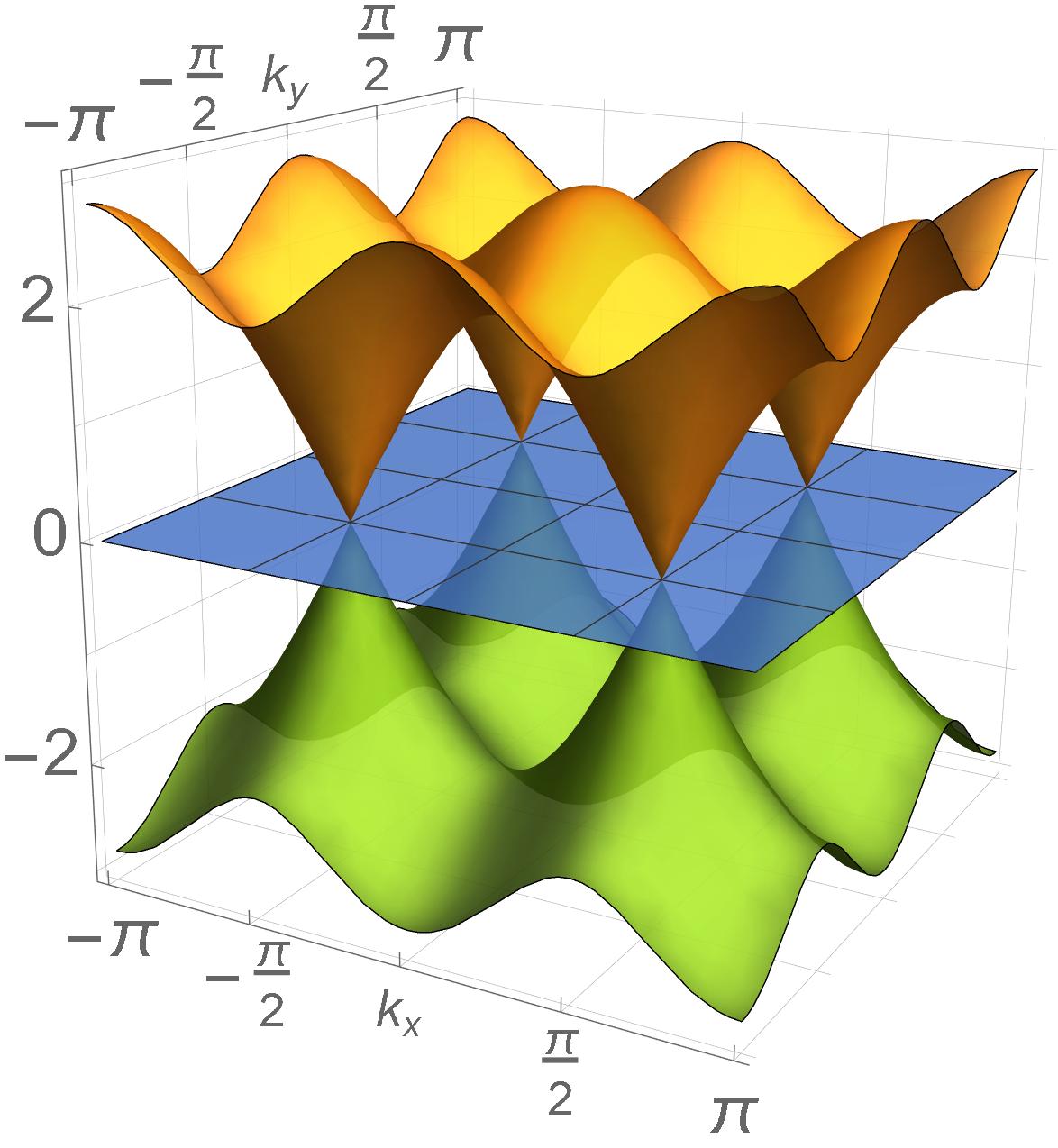}
\hspace{1mm}
\label{fig:DP0}}
\subfloat[][$t_{3}<\frac{t_{1}}{2}, t_{3}<\frac{t_{2}}{2}$]{\includegraphics[width=0.25\textwidth]{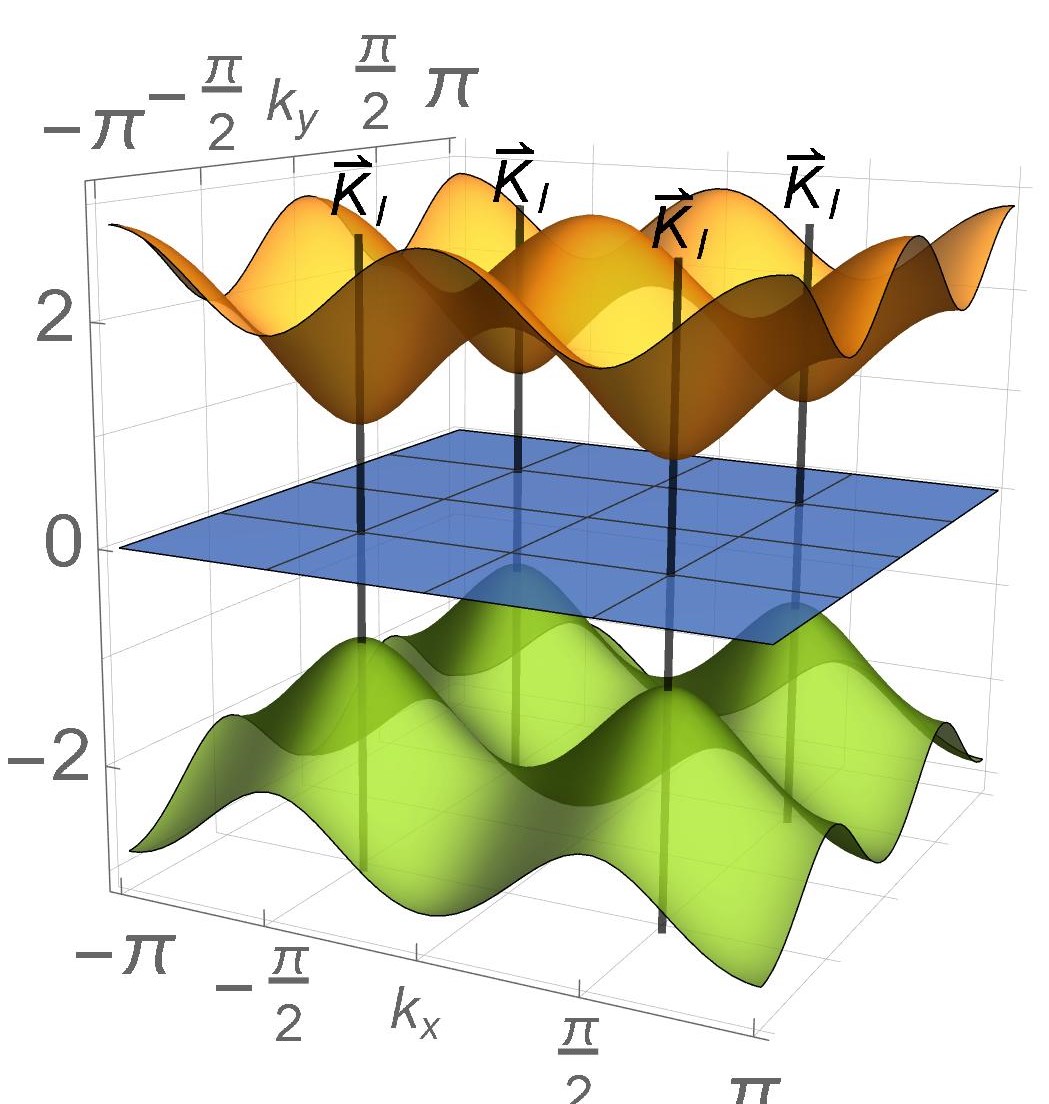}
\label{fig:DP1}}
\\
\subfloat[][$\frac{t_{2}}{2}<t_{3}<\frac{t_{1}}{2}$]{\includegraphics[width=0.25\textwidth]{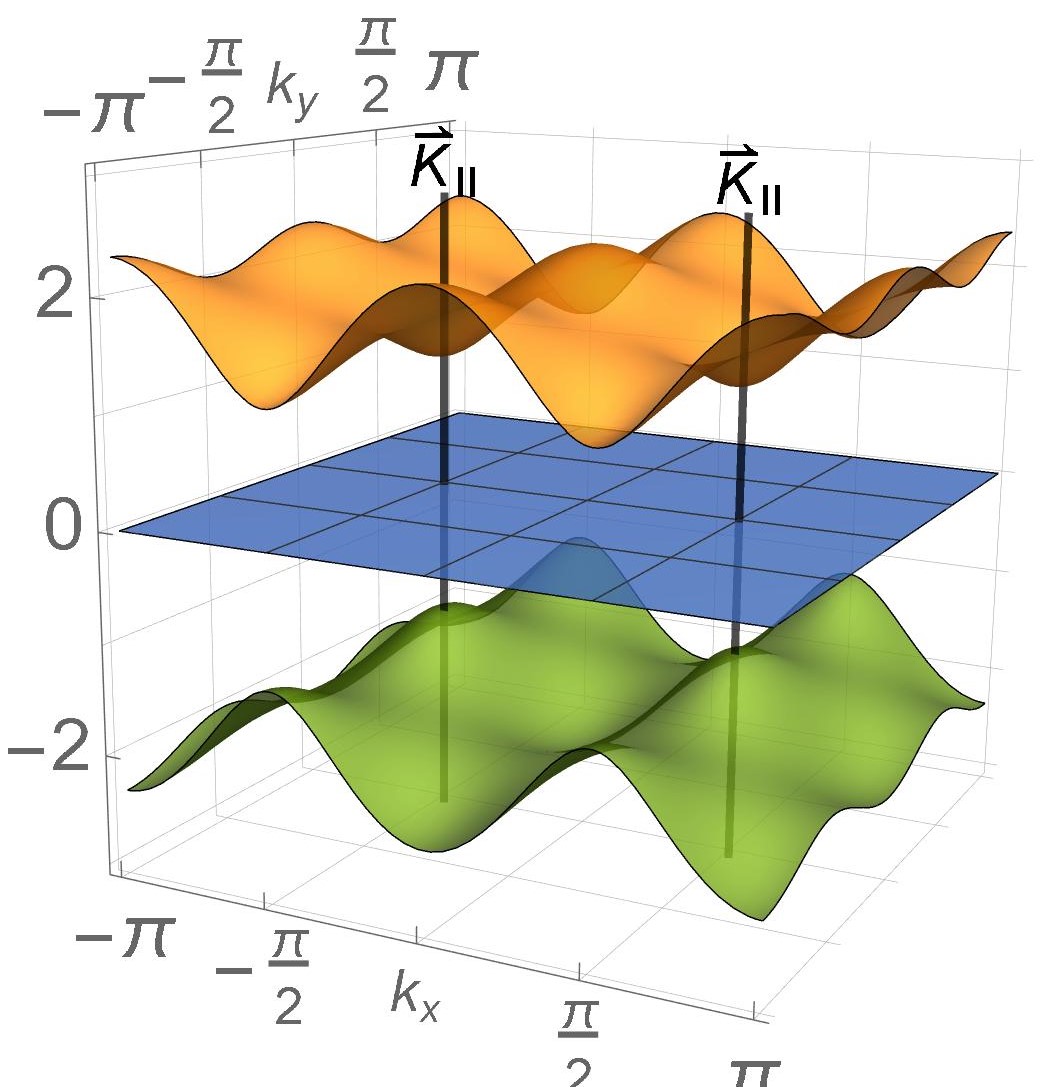}
\label{fig:DP2}}
\hspace{1mm}
\subfloat[][$\frac{t_{1}}{2}<t_{3}<\frac{t_{2}}{2}$]{\includegraphics[width=0.25\textwidth]{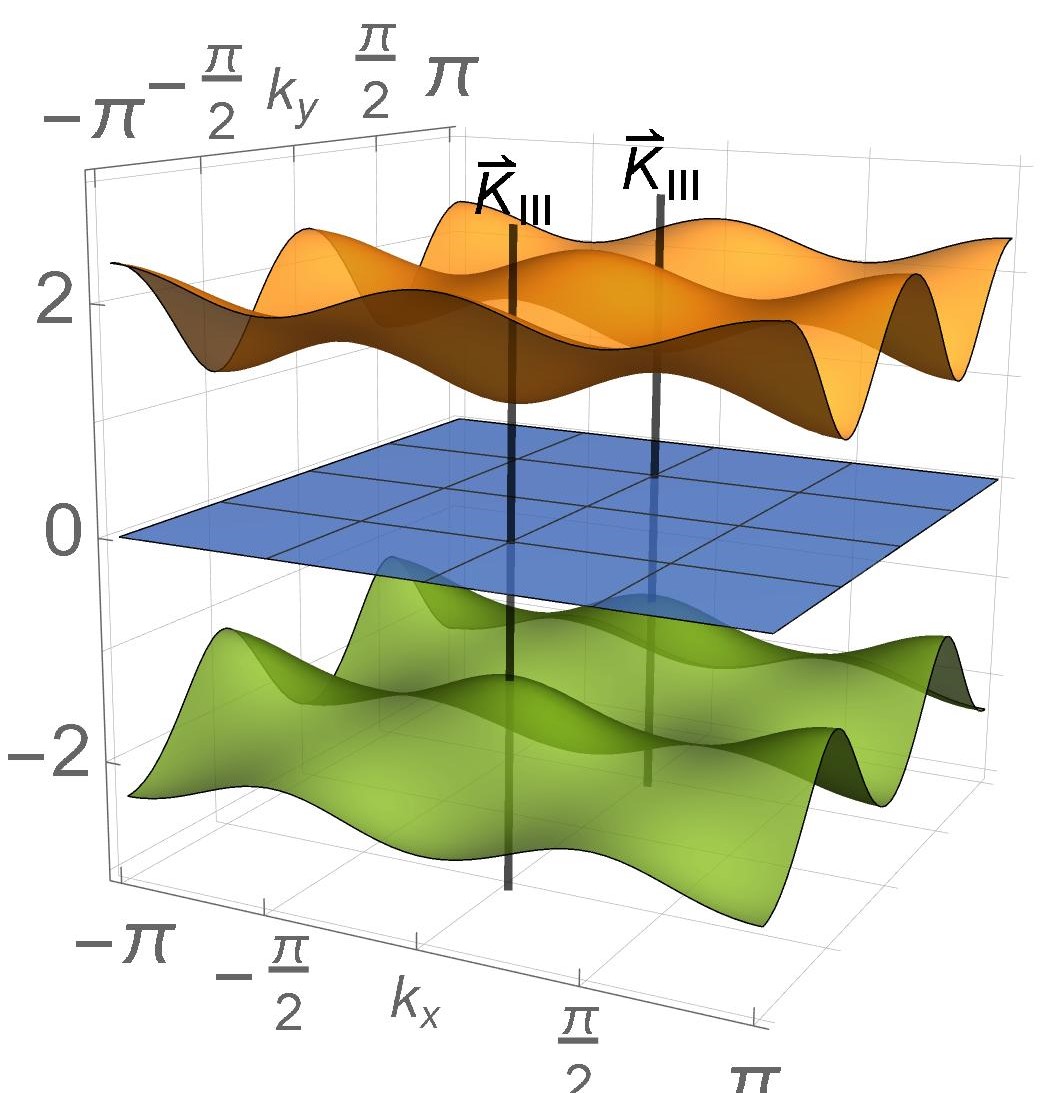}
\label{fig:DP3}}
\hspace{1mm}
\subfloat[][$t_{3}>\frac{t_{1}}{2}, t_{3}>\frac{t_{2}}{2}$]{\includegraphics[width=0.25\textwidth]{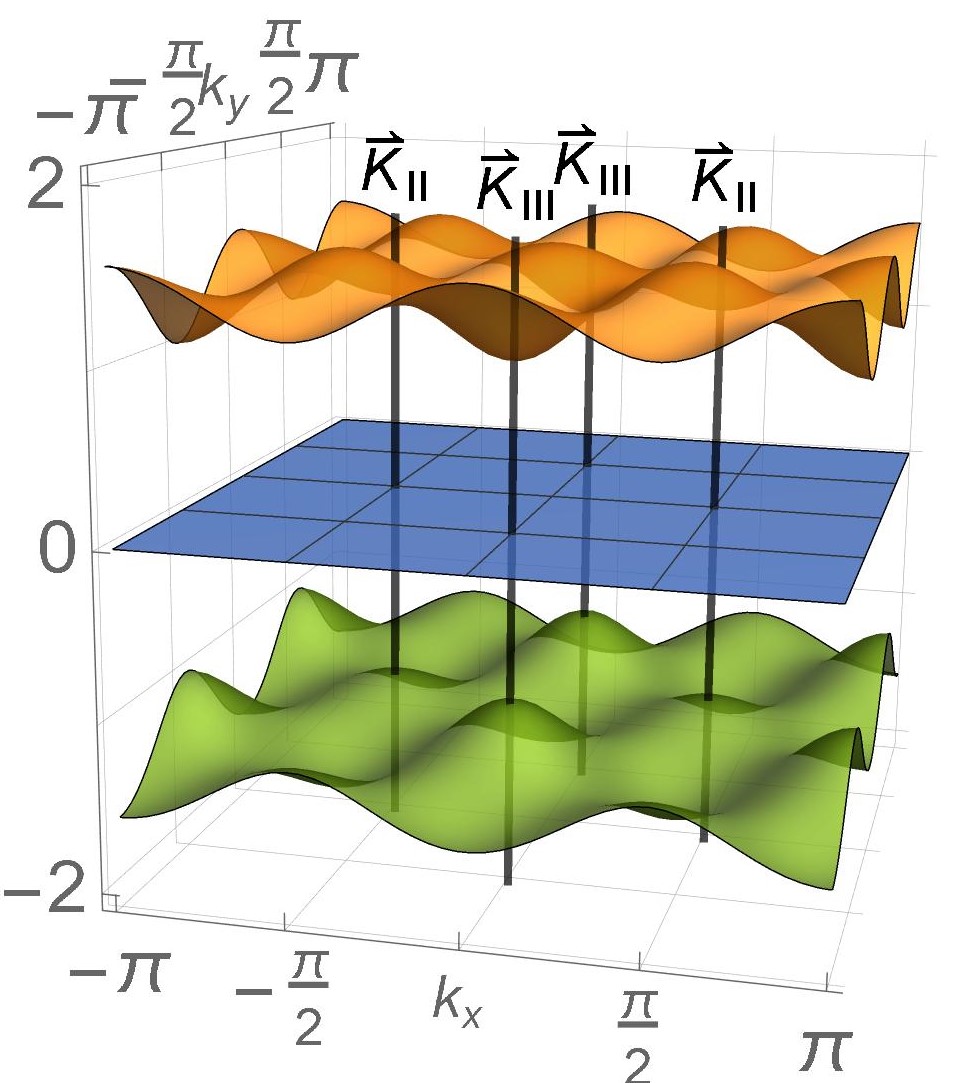}
\label{fig:DP4}}
\caption{Dispersion bands $w_{+}(\vec{k})$ (yellow-upper), $w_{-}(\vec{k})$ (green-lower), and $w_{0}(\vec{k})$ (blue-middle) for $\lambda=\frac{\pi}{2}$. The hopping parameters have been fixed as $\{t_{1}=t_{2}=1,t_{3}=0\}$ (a), $\{t_{1}=t_{2}=1,t_{3}=0.25\}$ (b), $\{t_{1}=1, t_{2}=0.6,t_{3}=0.4\}$ (c), $\{t_{1}=0.6, t_{2}=1,t_{3}=0.4\}$ (d), and $\{t_{1}=0.5, t_{2}=0.6,t_{3}=0.4\}$ (e), so that all the possible configurations in Tab.~\ref{tab:DiracP} are illustrated.}
\label{fig:Bands}
\end{figure}
The NNN interaction encoded in $t_3$ is usually considered to be much smaller than the NN interaction. In the literature, the phase $\lambda$ of the NNN interaction is frequently set to $\lambda=\frac{\pi}{2}$, see \cite{Gol11,Hwa21}, which is interpreted as the spin-orbital coupling. In this case, the dispersion bands $w_{j}(\vec{k})$ in~(\ref{characteristic}) acquire the simple form
\begin{equation}
w_{3}(\vec{k})=0, \quad w_{\pm}(\vec{k})=\pm\sqrt{t_{1}^{2}\cos^{2}\left(ak_{x}\right)+t_{2}^{2}\cos^{2}\left(ak_{y}\right)+4t_{3}^{2}\sin^{2}\left(ak_{x}\right)\sin^{2}\left(ak_{y}\right)}.
\label{band-pi2}
\end{equation}

To determine the Dirac valleys, low energy regions where an effective Dirac equation can be identified, we shall localize the minimum (maximum) point of $w_{+}(\vec{k})$ ($w_{-}(\vec{k})$) in the $\vec{k}$-space. This is particularly simple for $\lambda=\frac{\pi}{2}$.  From the points in the first Brillouin zone where $\vec{\nabla} w_+=\vec{0}$ is satisfied, the minimum is situated either in $\vec{K}_{\operatorname{I}} =\left(\frac{\pi}{2a},\frac{\pi}{2a}\right)$, $\vec{K}_{\operatorname{II}} =\left(\frac{\pi}{2a},0\right)$, or $\vec{K}_{\operatorname{III}} =\left(0,\frac{\pi}{2a}\right)$. Its actual position depends on the values of the parameters $t_{1,2,3}$.

For $t_{3}=0$, the three bands touch simultaneously at the point $\vec{K}_{\operatorname{I}}$, leading to a \textit{gapless} configuration and forming a Dirac cones around such a point. This is depicted in Fig.~\ref{fig:DP0}. 

For $t_{3}\neq 0$, the minimum of $w_+$ changes its position in dependence on the value of $t_3$. When $t_{3}<\frac{t_{1}}{2}$ and $t_{3}<\frac{t_{2}}{2}$, the three bands do no longer intersect,  the flat band is  \textit{gapped} symmetrically by $w_{\pm}$. The minimum (maximum) of $w_+$ ($w_-$) is still situated in  $\vec{K}_{\operatorname{I}}$  whereas there are saddle points in $\vec{K}_{\operatorname{II}}$ and $\vec{K}_{\operatorname{III}}$.  This is shown in Fig.~\ref{fig:DP1}. If $t_{3}>\frac{t_{2}}{2}$ and $t_{3}<\frac{t_{1}}{2}$, the minimum of $w_+$ sits at $\vec{K}_{\operatorname{II}}$, and $\vec{K}_{\operatorname{I}}$ and $\vec{K}_{\operatorname{III}}$ are the saddle points. For $t_{3}>\frac{t_{1}}{2}$ and $t_{3}<\frac{t_{2}}{2}$, the minimum is at $\vec{K}_{\operatorname{III}}$. Both cases are depicted in Fig.~\ref{fig:DP2} and Fig.~\ref{fig:DP3}, respectively. Finally, when $t_3>\frac{t_{1}}{2}$ and $t_3>\frac{t_{2}}{2}$, there are two minima situated at $\vec{K}_{\operatorname{III}}$ and $\vec{K}_{\operatorname{II}}$. This is the only configuration that admits two inequivalent Dirac valleys simultaneously, as can be seen in  Fig.~\ref{fig:DP4}.  All the possible configurations for Dirac valleys are summarized in Tab.~\ref{tab:DiracP}. It is worth saying that the configuration where  $t_{3}\ll t_{1,2}$ is the most relevant as the NNN interaction is usually much smaller than that of NN's. 

\begin{table}
\centering
\begin{tabular}{lll}
Condition & Location of Dirac valleys & Band gap\\
$t_{3}<\frac{t_{1}}{2}$, $t_{3}<\frac{t_{2}}{2}$ & $\vec{K}_{\operatorname{I}}=(\frac{\pi}{2a},\frac{\pi}{2a})$& $4t_3$\\
$\frac{t_{2}}{2}<t_{3}<\frac{t_{1}}{2}$ & $\vec{K}_{\operatorname{II}}=(\frac{\pi}{2a},0)$& $2t_2$ \\
$\frac{t_{1}}{2}<t_{3}<\frac{t_{2}}{2}$ & $\vec{K}_{\operatorname{III}}=\left(0,\frac{\pi}{2a}\right)$& $2t_1$ \\
$t_{3}>\frac{t_{1}}{2}$, $t_{3}>\frac{t_{2}}{2}$ & $\vec{K}_{\operatorname{II}}$, $\vec{K}_{\operatorname{III}}$
\end{tabular}
\caption{Location of the unequivalent Dirac valleys $\vec{K}_{\operatorname{j}}$ in the $\vec{k}$-space for $\lambda=\frac{\pi}{2}$, with $j=\operatorname{I},\operatorname{II},\operatorname{III}$, inside the first Brillouin zone. The band gap is determined by the difference between the minimum of $w_+$ and the maximum of $w_-$.}
\label{tab:DiracP}
\end{table}

Let us calculate the approximate form of the Hamiltonian $\mathcal{H}(\vec{k})$ in the vicinity of the three points. We denote the effective operator as $\mathcal{H}_{\operatorname{X}}(\vec{k})\equiv \mathcal{H}(\vec{K}_{X}+\vec{k})$, with $X=\operatorname{I,II,III}$, where $\vert\vec{k}\vert$ is considered small enough so that we can keep terms up to first-order in $\vec{k}$. The proper expansion of $\mathcal{H}(\vec{k})$ at the three Dirac points $\vec{K}_{X}$ can be conveniently written as
\begin{equation}
\begin{aligned}
& \mathcal{H}_{\operatorname{I}}(\vec{k})= 2a t_{1} k_{x} S_{1} + 2a t_{2} k_{y} S_{2} + 4 t_{3} S_{3} , \\
& \mathcal{H}_{\operatorname{II}}(\vec{k})= 2a t_{1} k_{x} S_{1} + 4a t_{3} k_{y} S_{3} - 2t_{2} S_{2} , \\
& \mathcal{H}_{\operatorname{III}}(\vec{k})= 4a t_{3} k_{x} S_{3} + 2a t_{2} k_{y} S_{2} - 2t_{1} S_{1}
, 
\end{aligned}
\label{Dirac-H1}
\end{equation}
where 
\begin{equation}
S_{1}=
\begin{pmatrix}
0 & 1 & 0 \\
1 & 0 & 0 \\
0 & 0 & 0
\end{pmatrix}
, \quad 
S_{2}=
\begin{pmatrix}
0 & 0 & 1 \\
0 & 0 & 0 \\
1 & 0 & 0
\end{pmatrix} 
, \quad 
S_{3}=
\begin{pmatrix}
0 & 0 & 0 \\
0 & 0 & -i \\
0 & i & 0
\end{pmatrix}
,
\label{matrix-spin1}
\end{equation}
stand for the three-dimensional spin 1 matrix representation. The latter satisfy the $su(2)$ commutation relations $[S_{p},S_{q}]=i\varepsilon_{pqr}S_{r}$, with $\varepsilon_{pqr}$ the three-dimensional anti-symmetric tensor. 

In the following two sections, we will discuss the spectral properties of the quasi-particles in the vicinity of the three Dirac points in the presence of a homogeneous and anti-symmetric magnetic field.

\section{Spin-1 particles in a homogeneous magnetic field}
\label{sec:spin1}
In order to study any potential confinement of electrons in the lattice, we introduce electromagnetic interactions into the low-energy Hamiltonian via minimal coupling. We particularly focus only on magnetic interactions, and no external electric fields are considered. This is achieved through the Peierls transformation~\cite{Pei29,Blo28},
\begin{equation}
\vec{k}\rightarrow -i\hbar\vec{\nabla}-\vec{A}(\vec{x}), \quad \vec{\nabla}=(\partial_{x},\partial_{y}), \quad \vec{A}(\vec{x})=(A_{x}(x,y),A_{y}(x,y)) ,
\label{peierls}
\end{equation}
where $\vec{A}$ is a two-dimensional vector potential related to the magnetic field $\vec{B}=\vec{\nabla}\times\vec{A}$.

For the sake of the analysis presented in the next sections, we present here the formulas for density currents. They can be obtained from the continuity equation $\partial_{t}\rho+\vec{\nabla}\cdot\boldsymbol{j}_{X}=0$  associated with the Hamiltonian $\mathcal{H}_X$ for $X=\operatorname{I}, \operatorname{II}, \operatorname{III}$. Here, $\rho=\boldsymbol{\Psi}^{\dagger}\boldsymbol{\Psi}$ is the probability density, with $\boldsymbol{\Psi} \equiv \boldsymbol{\Psi}(\vec{x})=(\psi_{A},\psi_{B},\psi_{C})^{T}$. Likewise, the current density is given by 
\begin{equation}
\begin{aligned}
& \boldsymbol{j}_{\operatorname{I}}=\left(4at_{1}\operatorname{Re}\psi_{A}^{*}\psi_{B},4at_{2}\operatorname{Re}\psi_{A}^{*}\psi_{C} \right) , \\
& \boldsymbol{j}_{\operatorname{II}}=\left(4at_{1}\operatorname{Re}\psi_{A}^{*}\psi_{B},8at_{3}\operatorname{Im}\psi_{B}^{*}\psi_{C} \right) , \\
& \boldsymbol{j}_{\operatorname{III}}=\left(8at_{3}\operatorname{Im}\psi_{B}^{*}\psi_{C},4at_{2}\operatorname{Re}\psi_{A}^{*}\psi_{C} \right) .
\end{aligned}
\label{current}
\end{equation}
The latter are general and hold valid even for discontinuous vector potentials $\vec{A}(x)$, as long as the latter do not have singularities.

Throughout this manuscript, we focus on vector potentials with translational invariance in the $\hat{y}$-axis, together with the Landau gauge $\vec{A}=A_{y}(x)\hat{y}$. This produces an external magnetic field perpendicular to the lattice of the form $\vec{B}=\partial_{x}A_{y}(x)\hat{z}$. 

The translational invariance allows us to conveniently rewrite the eigensolutions as
\begin{equation}
\boldsymbol{\Psi}(x,y)\rightarrow e^{ik_{2}y}(\psi_{A}(x),\psi_{B}(x),\psi_{C}(x))^{T}, \quad k_{2}\in\mathbb{R}, 
\label{psi-inv}
\end{equation}
where $k_{2}$ represents the continuous momentum on the $\hat{y}$-direction. This simplifies any derivative with respect to $y$ in the Hamiltonians as $-i\hbar\partial_{y}\rightarrow \hbar k_{2}$, leading to the effective Hamiltonians in the vicinity of the three Dirac points given as $H_X=\mathcal{H}_{X}\vert_{k_x\rightarrow -i\hbar\partial_x,\,k_y\rightarrow \hbar k_{2}-A_{y}(x)}$, with $X=\operatorname{I}, \operatorname{II}, \operatorname{III}$. 

Throughout this section, we are interested in a magnetic field homogeneous and perpendicular to the lattice; i.e.,
\begin{equation}
A_y(x)=B x, \quad \vec{B}=B\hat{z} ,
\label{HO}
\end{equation}
from which we obtain the corresponding Hamiltonians 
\begin{equation}
\begin{aligned}
& H_{\operatorname{I}}=-i\hbar (2at_{1})S_{1}\partial_{x}+2at_{2}(\hbar k_{2}-B x)S_{2}+4t_{3}S_{3} , \\
& H_{\operatorname{II}}=-i\hbar (2at_{1})S_{1}\partial_{x}+4at_{3}(\hbar k_{2}-B x)S_{3}-2t_{2}S_{2} , \\
& H_{\operatorname{III}}=-i\hbar (4at_{3})S_{3}\partial_{x}+2at_{2}(\hbar k_{2}-B x)S_{2}-2t_{1}S_{1} .
\end{aligned}
\end{equation}
for each Dirac point. The latter defines the eigenvalue equations $(H_X-E)(\psi_{A},\psi_{B},\psi_{C})^T=0$.

The dynamics at each of the three Dirac points can be decoupled such that two components of the wave function are given in terms of the third component. The straightforward calculations show that the latter component solves the Schr\"odinger equation of the stationary oscillator. Explicitly, the set of equations for the components of $\boldsymbol{\Psi}$ at each Dirac point can written as
\begin{equation}
-\psi_{p}''+\Omega^{2}\left(x-\frac{\hbar k_{2}}{B}\right)^{2}\psi_{p}=\mathcal{E} \psi_{p} ,
\label{psi-p}
\end{equation}
with $\psi_{p}''\equiv\frac{\partial^{2}\psi_{p}}{\partial x^{2}}$, together with
\begin{equation}
\psi_{q}=\nu_{1}\left(\frac{2at_{1}'\hbar E \psi_{p}'+8at_{2}'t_{3}'(\hbar k_{2}-Bx)\psi_{p}}{E^{2}-(4t_{3}')^{2}}\right) , \quad 
\psi_{r}=\nu_{2}\left(\frac{8at_{1}'t_{3}'\hbar\psi_{p}'+2at_{2}'E(\hbar k_{2}-Bx)\psi_{p}}{E^{2}-(4t_{3}')^{2}}\right) ,
\label{psi-q}
\end{equation}
where the parameters are given by
\begin{equation}
\Omega=\frac{t_{2}'}{t_{1}'}\frac{B}{\hbar} , \quad \mathcal{E}=\frac{E^{2}-(4t_{3}')^{2}}{\hbar^{2}(2at_{1}')^{2}}-\frac{4t_{2}'t_{3}'}{t_{1}'}\frac{B}{\hbar E} .
\end{equation}
The specific values for the indexes and the parameters for each Dirac point are shown in Tab.~\ref{tab:parameters}.

\begin{table}
\centering
\begin{tabular}{c|ccc|l}
Dirac                      & \multicolumn{3}{l|}{Indexes}                        & \multicolumn{1}{c}{\multirow{2}{*}{Parameters}} \\ \cline{2-4}
\multicolumn{1}{c|}{point} & \multicolumn{1}{l|}{p} & \multicolumn{1}{l|}{q} & r &                             \\ \hline
I                          & \multicolumn{1}{l|}{A} & \multicolumn{1}{l|}{B} & C & $t_{1}'\rightarrow t_{1}$, $t_{2}'\rightarrow t_{2}$, $t_{3}'\rightarrow t_{3}$, $\nu_{1}=-i$, $\nu_{2}=1$ \\
II                         & \multicolumn{1}{l|}{B} & \multicolumn{1}{l|}{A} & C & $t_{1}'\rightarrow t_{1}$, $t_{2}'\rightarrow 2t_{3}$, $t_{3}'\rightarrow \frac{t_{2}}{2}$, $\nu_{1}=-i$, $\nu_{2}=i$ \\
III                        & \multicolumn{1}{l|}{C} & \multicolumn{1}{l|}{B} & A & $t_{1}'\rightarrow 2t_{3}$, $t_{2}'\rightarrow t_{2}$, $t_{3}'\rightarrow \frac{t_{1}}{2}$, $\nu_{1}=-1$, $\nu_{2}=1$ 
\end{tabular}
\caption{Hopping parameters and indexes used in~\eqref{psi-p}-\eqref{psi-q} associated with each of the Dirac points in Tab.~\ref{tab:DiracP}.}
\label{tab:parameters}
\end{table}

Regardless of the Dirac point in question, the general solution for $\psi_{p}$ in~\eqref{psi-p} takes the form
\begin{equation}
\begin{aligned}
& \psi_{p}(x)=\ell_{1}F_{1}(x)+\ell_{2}F_{2}(x) , \quad \ell_{1},\ell_{2}\in\mathbb{R} , \\
& F_{1}(x)=e^{-\frac{z^{2}(x)}{2}}{}_{1}F_{1}\left(\frac{\vert\Omega\vert-\mathcal{E}}{4\vert\Omega\vert},\frac{1}{2},z^{2}(x)\right), \quad 
F_{2}(x)=e^{-\frac{z^{2}(x)}{2}} z(x) {}_{1}F_{1}\left(\frac{3\vert\Omega\vert-\mathcal{E}}{4\vert\Omega\vert},\frac{3}{2},z^{2}(x)\right) , \\
& z(x)=\sqrt{\vert\Omega\vert}\left( x-\frac{k_{2}}{\hbar B} \right) ,
\end{aligned}
\label{F1F2}
\end{equation}
where ${}_{1}F_{1}(a,b,z)$ stands for the confluent hypergeometric functions~\cite{Olv10}, $\ell_{1,2}$ are arbitrary real constants, and $\Omega$ and $\mathcal{E}$ take the corresponding values according to Tab.~\ref{tab:parameters}.

The solutions $\boldsymbol{\Psi}$ are square integrable provided that the hypergeometric functions are truncated to polynomials, otherwise, it grows faster than the Gaussian term for $\vert x\vert\rightarrow\infty$. The polynomial behavior is obtained if the first entry of the hypergeometric function is a negative integer or zero. Nevertheless, it is not possible that both hypergeometric functions in~\eqref{F1F2} fulfill this condition simultaneously for the same value of $E$, and we thus separate the discussion in two cases. For $\ell_{2}=0$, we impose that $1-\mathcal{E}/\vert\Omega\vert=-4n$, whereas for $\ell_{1}=0$ we set $3-\mathcal{E}/\vert\Omega\vert=-4n$, for $n=0,1,\ldots$. Those conditions reduce $F_{1}(x)$ and $F_{2}(x)$ to a Gaussian function times an even $\texttt{H}_{2n}(z)$ and odd $\texttt{H}_{2n+1}(z)$ Hermite polynomials\cite{Olv10}, respectively, so that we obtain the simpler relation that leads to quantization of $\mathcal{E}$, 
\begin{equation}
1-\frac{\mathcal{E}}{\vert \Omega \vert}=-2n, \quad n=0,1,\dots.
\label{quatization}
\end{equation}

Clearly, the overall behavior of the eigensolutions $\boldsymbol{\Psi}$ is the same for every Dirac point, and thus, it is just necessary to find the dynamics at one point. The dynamics for the other points can be found straightforwardly by using the permutations listed in Tab.~\ref{tab:parameters}. For this reason, and without loss of generality, we focus on the Dirac point $\vec{K}_{\operatorname{I}}$. Remark that this argument is valid as long as we stay at $\lambda=\pi/2$, as the latter permutations are not necessarily valid for other values of $\lambda$. 

\subsection{Cardano's formula, Landau levels, and electron confinement}
\label{sec:landau}
Here we focus on the Dirac point $\vec{K}_{\operatorname{I}}$. The results obtained throughout this section can be translated to the other points by using the permutations in Tab.~\ref{tab:parameters}. For convenience and to simplify our notation, we reparametrize the hopping parameters as
\begin{equation}
v_{1}=2at_{1} , \quad v_{2}=2at_{2} , \quad m=4t_{3} ,
\label{para}
\end{equation}
where $v_{1}$ and $v_{2}$ play the role of the anisotropic \textit{Fermi velocity} on the $\hat{x}$- and $\hat{y}$-direction, respectively. Likewise, $m$ plays the role of the \textit{mass term} in the Dirac equation, which is responsible for the gap in the band structure, which is constrained as $m\in\left[0,min\left(\frac{v_{1}}{a},\frac{v_{2}}{a}\right)\right]$. 

In this form, following~\eqref{psi-p}, we get the constants
\begin{equation}
\vert \Omega\vert = \frac{v_{2}}{v_{1}}\frac{\vert B \vert}{\hbar} , \quad \mathcal{E}=-\frac{v_{2}m}{v_{1}}\frac{B}{\hbar E}+\frac{E^{2}-m^{2}}{\hbar^{2}v_{1}^{2}} ,
\label{Omega-E}
\end{equation}
from which the eigensolutions and eigenvalues can be determined explicitly. Before doing so, it is worth noticing that, for $E=\pm m=\pm 4t_{3}$, Eqs.~\eqref{psi-q} become proportional to each other, and thus the corresponding eigensolution components shall be determined by other means; i.e., solutions can be extracted by substituting $E=\pm m$ in the stationary equation for $H_{\operatorname{I}}$. Nevertheless, this case usually leads to non-finite-norm solutions, as discussed below.

From $\mathcal{E}$ and $\Omega$ in~\eqref{Omega-E}, together with the finite-norm condition~\eqref{quatization}, one realizes that the eigenvalues are determined from the cubic polynomial equation
\begin{equation}
E^{3}-\left( m^{2}+\hbar \vert B\vert v_{1}v_{2}(2n+1) \right)E-\hbar B m v_{1}v_{2}=0 ,
\label{E-cubic}
\end{equation}
whereas the component $\psi_{A}$ can de computed explicitly as
\begin{equation}
\psi_{A;n}(x)\propto e^{-\frac{z^{2}(x)}{2}}\texttt{H}_{n}(z(x)) , \quad z(x)=\sqrt{\frac{v_{2}}{v_{1}}\frac{\vert B\vert}{\hbar}}\left( x-\frac{\hbar k_{2}}{B} \right) \quad n=0,1,\ldots ,
\label{osc-cubic}
\end{equation}
which are the stationary oscillator eigensolutions.

Clearly, the transverse momentum $k_{2}$ is not involved in the eqaution for the eigenvalues $E$, but it does produce a displacement on the eigensolutions.

It is worth remarking that $n=0$ leads immediately to the solution $E=-m$ for $B>0$, and $E=m$ for $B<0$; the two remaining solutions can be obtained by reducing the order of the equation. After some reordering of the energies, we have
\begin{equation}
\begin{array}{llll}
B>0: \quad & E_{0}^{(1)}=\frac{m+\sqrt{m^{2}+4\hbar\vert B \vert v_{1}v_{2}}}{2}  \quad & E_{0}^{(2)}=-m , \quad  &E_{0}^{(3)}=\frac{m-\sqrt{m^{2}+4\hbar\vert B \vert v_{1}v_{2}}}{2} , \\
 B<0:  &E_{0}^{(1)}=\frac{-m+\sqrt{m^{2}+4\hbar\vert B \vert v_{1}v_{2}}}{2} , & E_{0}^{(2)}=m , & E_{0}^{(3)}=\frac{-m-\sqrt{m^{2}+4\hbar\vert B \vert v_{1}v_{2}}}{2} .
\label{osc-ground}
\end{array}
\end{equation}
In the latter, we have $E_{0}^{(1)}>E_{0}^{(2)}>E_{0}^{(3)}$ for $\frac{\hbar \vert B\vert}{2} > \frac{m^{2}}{v_{1}v_{2}}$. Likewise, we get $E_{0}^{(1)}>E_{0}^{(3)}>E_{0}^{(2)}$ for $\frac{\hbar \vert B\vert}{2} < \frac{m^{2}}{v_{1}v_{2}}$. 

Although the component $\psi_{A}$ has a finite-norm for $B>0$ ($B<0$) and $E=-m$ ($E=m$), the remaining components $\psi_{B}$ and $\psi_{C}$ diverge asymptotically. Thus, the eigenvalue $E=-m$ ($E=m$) does not belong to the discrete spectrum of $H_{\operatorname{I}}$.

Eq.~\eqref{E-cubic} can be solved for arbitrary $n$ by using the well-known Cardano's solutions~(see App.~\ref{sec:cardano}). For the cubic equation like $z^{3}+pz+q=0$, the sign of discriminant $\Delta=4p^{3}+27q$ tells us whether the zeros are real or complex~\cite{Olv10}. It is known that for $\Delta<0$ all zeros are real and distinct. Particularly, the discriminant $\Delta_{n}$ of the cubic equation~\eqref{E-cubic} is a decreasing function with respect to $n$ when $B>0$, which implies that $\Delta_{n}<\Delta_{m}$ for all $n>m$. In this form, $\Delta_{0}$ establishes an upper bound for $\Delta_{n}$. One can easily show that $\Delta_{0}<0$ for $B>0$, and thus $\Delta_{n}<0$ for $n=0,1,\ldots$. 
We can thus ensure the existence of three different and real roots of $E$ in~\eqref{E-cubic} for each $n$ and $B\neq 0$. This is in agareement with the fact that $H_{\operatorname{I}}$ is a Hermitian operator. We denote the real eigenvalues by $E_{n}^{(j)}$, see App.~\ref{sec:cardano},
\begin{equation}
\begin{aligned}
& E_{n}^{(j)}=2R_{n} \cos\left( \frac{\theta_{n}-2(j-1)\pi}{3} \right) ,  \\ 
& R_{n}=\sqrt{\frac{m^{2}+\hbar \vert B\vert v_{1}v_{2}(2n+1)}{3}} , \quad \theta_{n}=\arccos\left(\frac{\hbar B v_{1}v_{2}m}{2R_{n}^{3}}\right) , 
\end{aligned}
\label{E-cardano}
\end{equation}
for $j=1,2,3$ and $n=0,1,\ldots$. 

Remark that, for $n=0$, only two of the three eigenvalues lead to finite norm solutions. Thus, the discrete spectrum of the Hamiltonian in question becomes
\begin{equation}
\sigma(H_{\operatorname{I}}) = \{E_{n}^{(1)}\}_{n=0}^{\infty}\cup\{E_{n+1}^{(2)}\}_{n=0}^{\infty}\cup\{E_{n}^{(3)}\}_{n=0}^{\infty} .
\label{Landau-Lieb}
\end{equation}
Further information can be extracted by analyzing the behavior of the eigenvalues for $n\rightarrow\infty$, from which we see that $\theta_{n}\vert_{n\rightarrow\infty}\rightarrow\pi/2$. Since $n$ grows indefinitely, it is always possible to find large values of $n$ such that $R_{n}^{3}\gg \hbar B v_{1}v_{2}m$, where the eigenvalues behave asymptotically as $E_{n}^{(1)}\approx 2\cos(\pi/6)R_{n}$, $E_{n}^{(2)}\approx 0$, and $E_{n}^{(3)}\approx -2\cos(\pi/6)R_{n}$. That is, the sequence $E_{n}^{(2)}$ has a lower bound at\footnote{Recall that $E_{0}^{(2)}$ does not produce regular eigensolutions, and it is thus discarded.} $E_{1}^{(2)}$ and an upper bound at zero, whereas the sequences $E_{n}^{(1)}$ and $E_{n}^{(3)}$ are only bounded from below and above, respectively. 

\begin{figure}
\centering
\subfloat[][]{\includegraphics[width=0.3\textwidth]{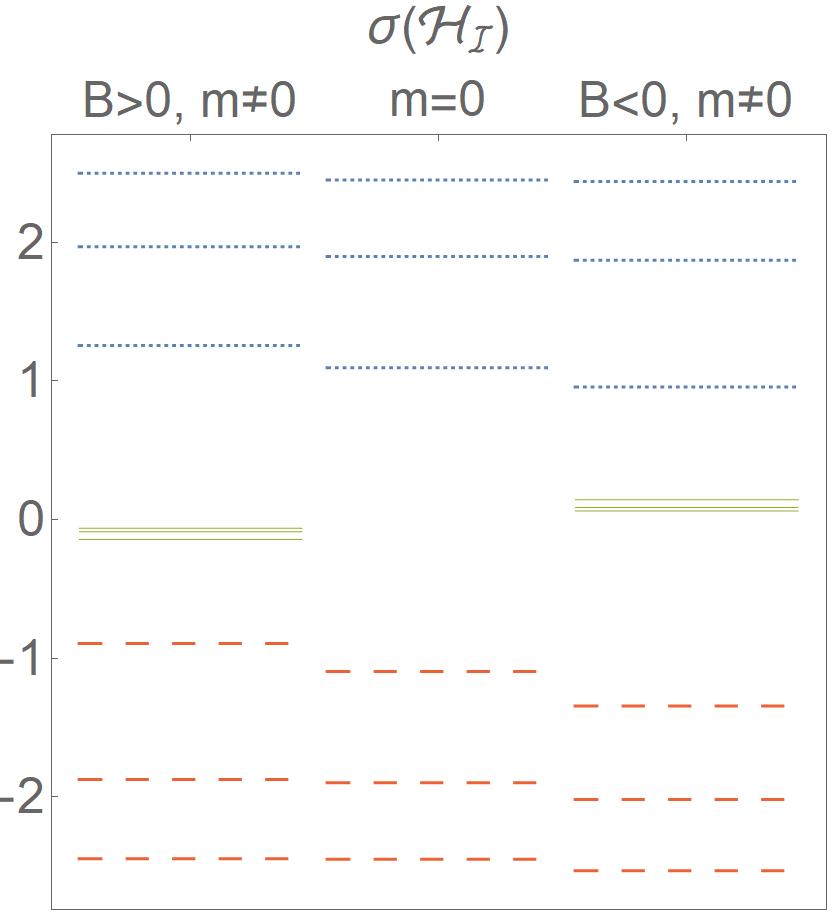}
\label{fig:spec-1}}
\subfloat[][$B>0$]{\includegraphics[width=0.28\textwidth]{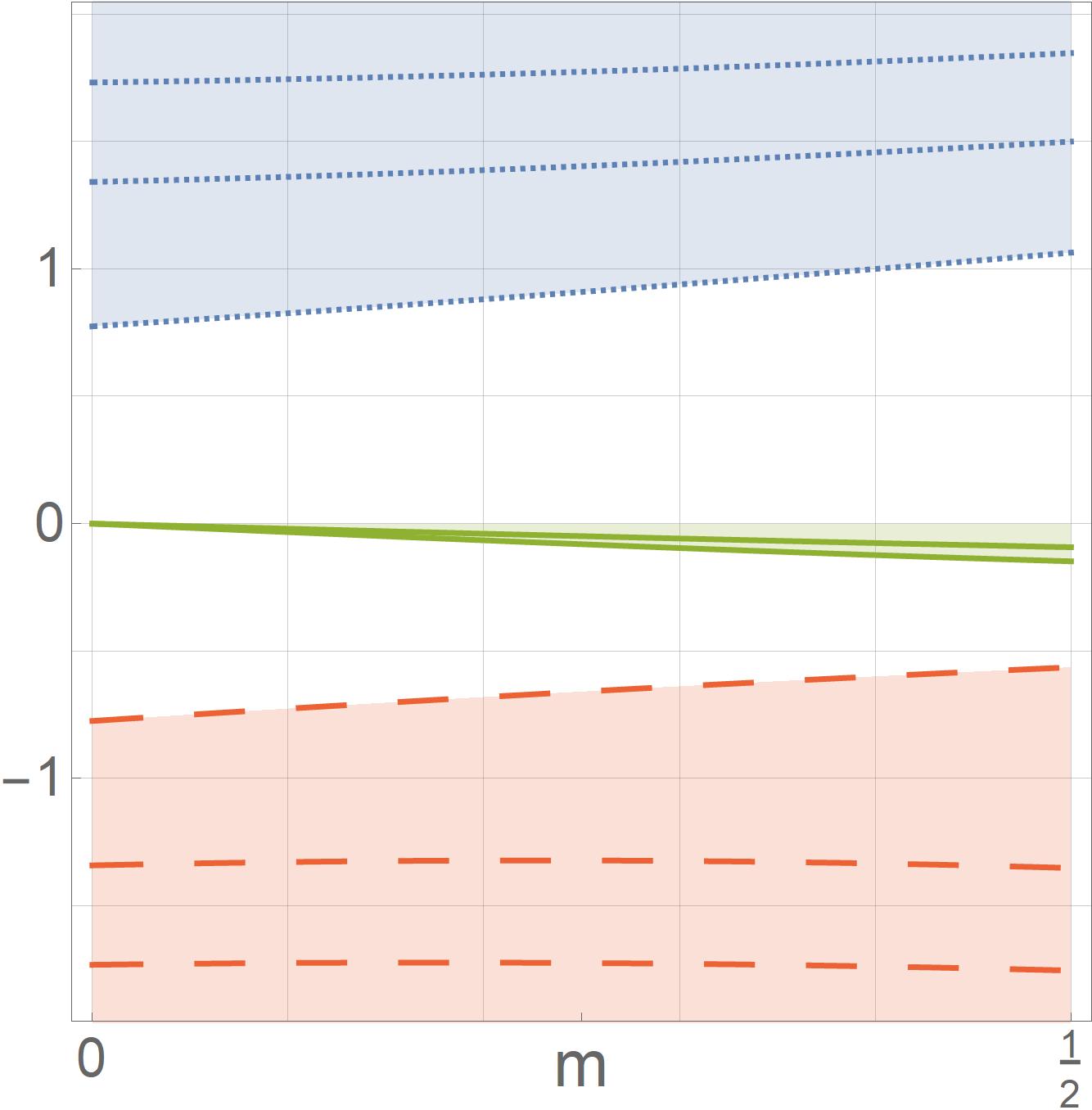}
\label{fig:spec-2}}
\subfloat[][$B<0$]{\includegraphics[width=0.28\textwidth]{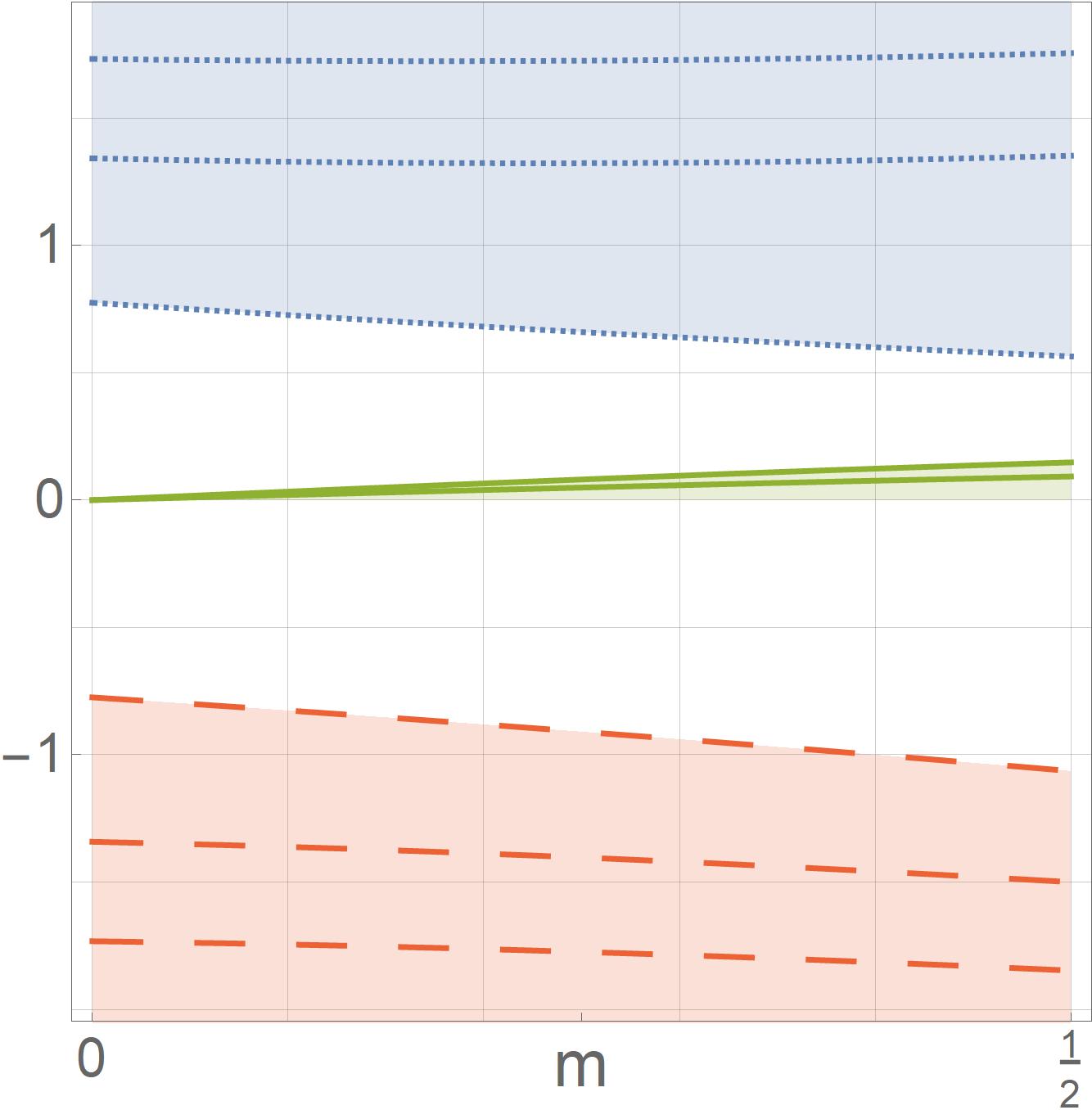}
\label{fig:spec-3}}
\caption{(a) Eigenvalues $E_{n}^{(j)}$ for $j=1$ (blue-dotted), $j=2$ (green-solid), and $j=3$ (red-dashed). The magnetic field has been fixed as $B=1.2$ (left levels) and $B=1.2$ (right levels), together with $m=0.45$ in all cases. For $m=0$, the sign of $B$ does not modify the eigenvalues. (b)-(c) Eigenvalues $E_{n}^{(j)}$ as a function of the next-nearest neighbor hopping parameter $m\in[0,\frac{1}{2}]$ for $B=1.2$ (b) and $B=-1.2$ (c). The shadowed area depicts the allowed regions in which eigenvalues of each sequence are embedded. The rest of parameters have been fix to $\hbar=v_{1}=v_{2}=1$.}
	\label{fig:osc-spec1}
\end{figure}

Furthermore, for $m\neq 0$ and $2n+1\gg (\hbar v_{1}v_{2}\vert B\vert)^{-1}m^{2}$, the top and bottom sequences of eigenvalues can be approximate to
\begin{equation}
E_{n}^{(1)}\approx \sqrt{(2n+1)\hbar \vert B\vert v_{1}v_{2}} , \quad E_{n}^{(3)}\approx -\sqrt{(2n+1)\hbar \vert B\vert v_{1}v_{2}} .
\end{equation}
In other words, for large enough values of $n$, the eigenvalue sequences approach to those corresponding to the gapless case $m=0$. Such a behavior can be seen in Fig.~\ref{fig:spec-1} where the eigenvalues $E_{n}^{(j)}$ are depicted for $B>0$ and $B<0$, as well as $m\neq 0$ and $m=0$. The infinite sequence of eigenvalues $E_{n}^{(2)}$ pile up below $E=0$ for $B>0$ and above $E=0$ for $B<0$.  It is worth to remark that, for $m=0$ (no NNN interaction), the whole infinite sequence of eigenvalues $E_{n}^{(2)}$ degenerates into a single eigenvalue $E^{(2)}=0$, which corresponds to the flat band of the free particle case. For $m\neq 0$, an infinite sequence of Landau levels are generated around the flat band. 

In summary, we have $E_{n}^{(1)}\in[E_{0}^{(1)},\infty)$, $E_{n}^{(3)}\in(-\infty,E_{0}^{(3)}]$, and $E_{n}^{(2)}\in[E_{1}^{(2)},0)$. Thus, the eigenvalues from different sequences never overlap, and the existence of level crossing is ruled out. To illustrate the latter, we depict in Fig.~\ref{fig:spec-1}-\ref{fig:spec-3} the behavior of $E_{n;j}$ as a function of $m$, for $j=0,1,2$ and several values of $n$. 

Following~\eqref{psi-q} and~\eqref{osc-cubic}, we compute the remaining components of the eigensolutions. The straightforward calculations lead to
\begin{equation}
\boldsymbol{\Psi}_{n}^{(j)}(x)=\mathcal{N}_{n}
\begin{pmatrix}
\left((E_{n}^{(j)})^{2}-m^{2}\right)\psi_{A;n}(x) \\
i\sqrt{h \vert B\vert v_{1}v_{2}} \left[ \left( E_{n}^{(j)}+sign(B) m \right) z(x) \psi_{A;n}(x) - 2n E_{n}^{(j)}\psi_{A;n-1}(x) \right] \\
-\sqrt{h \vert B\vert v_{1}v_{2}} \left[ \left( sign(B)E_{n}^{(j)}+m\right) z(x) \psi_{A;n}(x) - 2n m \psi_{A;n-1}(x) \right]
\end{pmatrix}
, 
\label{osc-sols}
\end{equation}
where $\psi_{A;n}$ are the harmonic oscillator eigensolutions given in~\eqref{osc-cubic} and $sign(x)=x/\vert x\vert$. The latter is valid for $j=1,3$ with $n=0,1\ldots$, and $j=2$ with $n=1,2\ldots$.
The normalization factor follows straightforwardly from the well-documented integrals involving Hermite polynomials. For instance, see Sec.~2.20.17 in~\cite{Pru86}. This leads to
\begin{multline}
\vert\mathcal{N}_{n}^{(j)}\vert=\left(2^{n}n!\sqrt{\frac{\hbar\pi}{\vert B\vert}\frac{v_{1}}{v_{2}}}\right)^{-\frac{1}{2}} 
\left[ \left((E_{n}^{(j)})^{2}-m^{2}\right)^{2} + \right. \\
\left. \hbar \vert B\vert v_{1}v_{2} \left( 
\left( E_{n}^{(j)})^{2}+m^{2}\right) (2n+1) + 2m E_{n}^{(j)}sign(B)   \right)\right]^{-\frac{1}{2}} .
\end{multline}

The probability current $\boldsymbol{j}_{\operatorname{I}}\equiv\boldsymbol{j}_{n}^{(j)}=(j_{x;n}^{(j)},j_{y;n}^{(j)})$ can be directly computed from~\eqref{current} and~\eqref{osc-sols}. This leads to a null probability current across the $\hat{x}$ direction for any allowed value of $j$ and $n=0,$, $j_{x;n}^{(j)}=0$. In turn, the probability current along the $\hat{y}$ direction is non-null, $j_{y;n}^{(j)}(x)=2v_{2}\psi_{A;n}^{(j)}\psi_{C;n}^{(j)}$. 

Despite $j_{y;n}^{(j)}$ being non-null, one can verify that no net current flows in the $\hat{y}$ direction. That is, the net current $J_{y;n}^{(j)}:=\int_{\mathbb{R}}dx j_{y;n}^{(j)}=0$ vanishes for any allowed $n$ and $j$, a property easily verifiable from the orthogonality of the oscillator solutions $\psi_{A;n}$. This means that local currents across $x\in\mathbb{R}$ compensate each other so that their net effect is null. This agrees with the concept of group velocity~\cite{Gho08} (a detailed proof for Bloch waves is provided in App.~E of~\cite{Ash76}), 
\begin{equation}
v_{y}(k_{2}):=\int_{\mathbb{R}}dx \, j_{y}(x;k_{2})\equiv\frac{\partial E(k_{2})}{\partial k_{2}},
\label{vel-group}
\end{equation}
which is nothing but the net current $J_{y}(k_{2})$. 
For Landau levels, the group velocity in the $\hat{y}$ direction vanishes since $E_{n}^{(j)}$ is independent of $k_{2}$, or alternatively because $J_{y}=0$.
This lack of transport of Dirac fermions is in accordance with the classical picture where the electrons move along circular trajectories the zero total flux.

It is worth to remark that the dynamics for the other Dirac points $\vec{K}_{\operatorname{II}}$ and $\vec{K}_{\operatorname{III}}$ can be recovered from the general results here presented by using the permutations shown in Tab.~\ref{tab:parameters}.

\section{Snake states in anti-symmetric magnetic field }
\label{sec:snake}
In Sec.~\ref{sec:landau}, we already showed that no net current is produced for Dirac fermions subjected to a homogeneous (symmetric) magnetic field. Thus, following~\cite{Mue92}, we consider the effects of an anti-symmetric magnetic field acting on the lattice. This can be easily achieved by considering a magnetic field that changes the sign on the two half-planes, i.e.,
\begin{equation}
\vec{B}=
\begin{cases}
B\hat{z} \quad & x>0 \\
-B\hat{z} \quad & x<0
\end{cases} , \quad
\vec{A}=\vert x\vert \hat{y}.
\label{anti-magn}
\end{equation}
In the classical case, there is circular motion of electrons that are far away from the interface. On the interface, the electrons follow snake-like trajectories as the Lorenz force acts in opposite directions on the two sides of the interface. The quantum system of Dirac fermions in graphene in presence of antisymmetric magnetic field was studied in \cite{Oro08}. The quantum analogue of the snake states was found in the form of confined states whose probability current along the interface was non-vanishing. Let us analyze this situation for Dirac fermions in the Lieb lattice.

We focus here on the most relevant case where Dirac cone is situated at $\vec{K}_{\operatorname{I}}$, i.e. $t_3<<t_{1,2}$. Moreover, we already know the general solution for the component $\psi_{A}$ in terms of the confluent hypergeometric function for a homogeneous external field. Using the latter, we construct the solutions for this problem as
\begin{equation}
\begin{aligned}
&\psi_{A}^{+}(x)=\ell_{1}^{+}F^{+}_{1}(x)+\ell_{2}^{+}F^{+}_{2}(x) , \quad x>0 , \\
&\psi_{A}^{-}(x)=\ell_{1}^{-}F^{-}_{1}(x)+\ell_{2}^{-}F^{-}_{2}(x) , \quad x>0 , 
\end{aligned}
\label{snake-psiA}
\end{equation}
with $F_{j}^{\pm}(x)=F_{j}(x)\vert_{B\rightarrow\pm\vert B\vert}$, and $F_{j}(x)$ given in~\eqref{F1F2} for $j=1,2$. The remaining solution components $\psi_{B}$ and $\psi_{C}$ are determined from~\eqref{psi-q}. 

The magnetic field is discontinuous at $x=0$, and thus the corresponding solutions must fulfill the boundary conditions $\psi_{A}(-\delta)=\psi_{A}(\delta)$ and $\psi_{B}(-\delta)=\psi_{B}(\delta)$ for $\delta\rightarrow 0$. This implies continuity of only two components of the eigensolution $\boldsymbol{\Psi}$. Since we are interested in electron confinement under the influence of $\vec{B}$, we seek for solutions such that $\lim_{x\rightarrow\pm\infty}\boldsymbol{\Psi}\rightarrow \vec{0}$; that is, $\boldsymbol{\Psi}\in L^{2}(\mathbb{R})\otimes\mathbb{C}^{3}$. 

We thus fix $\ell_{1,2}^{\pm}$ to fulfill the required boundary conditions. First, we require that the wave functions are asymptotically vanishing. This can be done with the aid of the asymptotic behavior for the confluent hypergeometric function, ${}_{1}F_{1}(\alpha,\beta;y)\vert_{y\rightarrow\infty} \sim \frac{\Gamma(\beta)}{\Gamma(\alpha)}e^{y}y^{\alpha-\beta}$. Implementing the latter into~\eqref{snake-psiA} and imposing the vanishing asymptotic behavior, $\psi^{\pm}_{A}\vert_{x\rightarrow\pm\infty}\rightarrow 0$, we find the relation $\ell_{2}^{\pm}=-2\ell_{1}^{\pm}\Gamma\left(\frac{3\vert\Omega\vert-\mathcal{E}_{\pm}}{4\vert\Omega\vert}\right)/\Gamma\left(\frac{\vert\Omega\vert-\mathcal{E}_{\pm}}{4\vert\Omega\vert}\right)$, with $\mathcal{E}_{\pm}:=\mathcal{E}\vert_{B\rightarrow\pm\vert B\vert}$. 
One may note that, after substituting the latter coefficients in~\eqref{snake-psiA}, the \textit{parabolic cylinder function} $D_{\nu}(y)$ appears quite straightforwardly (see Eq.~9.240 in~\cite{Gra07}). We thus get, up to a proportionality factor absorbed by $\ell_{1}^{\pm}$, the solutions
\begin{equation}
\begin{aligned}
& \psi_{A}^{\pm}(x)= \ell_{1}^{\pm} D_{\nu_{\pm}}(\pm\sqrt{2}z_{\pm}(x)) , \quad  
z_{\pm}(x)= \sqrt{\frac{v_{2}}{v_{1}}\frac{\vert B\vert}{\hbar}}\left(x\mp\frac{\hbar k_{2}}{\vert B\vert} \right), \\
& \nu_{\pm}=\frac{E^{3}-\left(m^{2}+\hbar\vert B\vert v_{1}v_{2}\right)E \mp \hbar\vert B\vert v_{1}v_{2}m}{2\hbar \vert B \vert v_{1}v_{2} E} .
\end{aligned}
\end{equation}

The boundary conditions at $x=0$ fix the relation between $\ell_{1}^{\pm}$ as they shall solve a homogeneous system of two equations for $\ell_{1}^{+}$ and $\ell_{1}^{-}$. It admits nontrivial solutions as long as the secular equation $\mathcal{S}(k_{2},E)=0$ holds, where
\begin{equation}
\mathcal{S}(k_{2},E)=-z_{0}D_{\nu_{-}}(z_{0}) D_{\nu_{+}}(z_{0})+D_{\nu{-}}(z_{0})D_{\nu_{+}+1}(z_{0})+D_{\nu_{+}}(z_{0})D_{\nu_{-}+1}(z_{0}),
\label{secular}
\end{equation}
and $z_{0}=-\sqrt{\frac{2\hbar v_{2}}{\vert B\vert v_{1}}}k_{2}$.  That is, we find the energies $E$ such that $\mathcal{S}(k_{2},E)\vert_{E=\epsilon_{n}^{(\kappa)}(k_{2})}=0$, from which we determine the dispersion bands $\epsilon_{n}^{(\kappa)}(k_{2})$. The indexes $n$ and $\kappa$ have been introduced in analogy to the Landau levels~\eqref{E-cardano}, as the secular equation leads to a countable set of dispersion bands. The meaning of such indexes is explained below.

The secular equation defines a transcendental equation that cannot be solved by exact means. Despite this, one may note that $\nu_{\pm}\vert_{m=0}=-\frac{E^{2}}{2\hbar \vert B\vert v_{1}v_{2}}-\frac{1}{2}$, so that the secular equation is invariant under the change $E\rightarrow-E$ and $B\rightarrow-B$. In other words, for $m=0$, we expect a band structure symmetric with respect to $E=0$ and independent of the direction of the magnetic field. 

\begin{figure}
\centering
\subfloat[][$m=0$]{\includegraphics[width=0.6\textwidth]{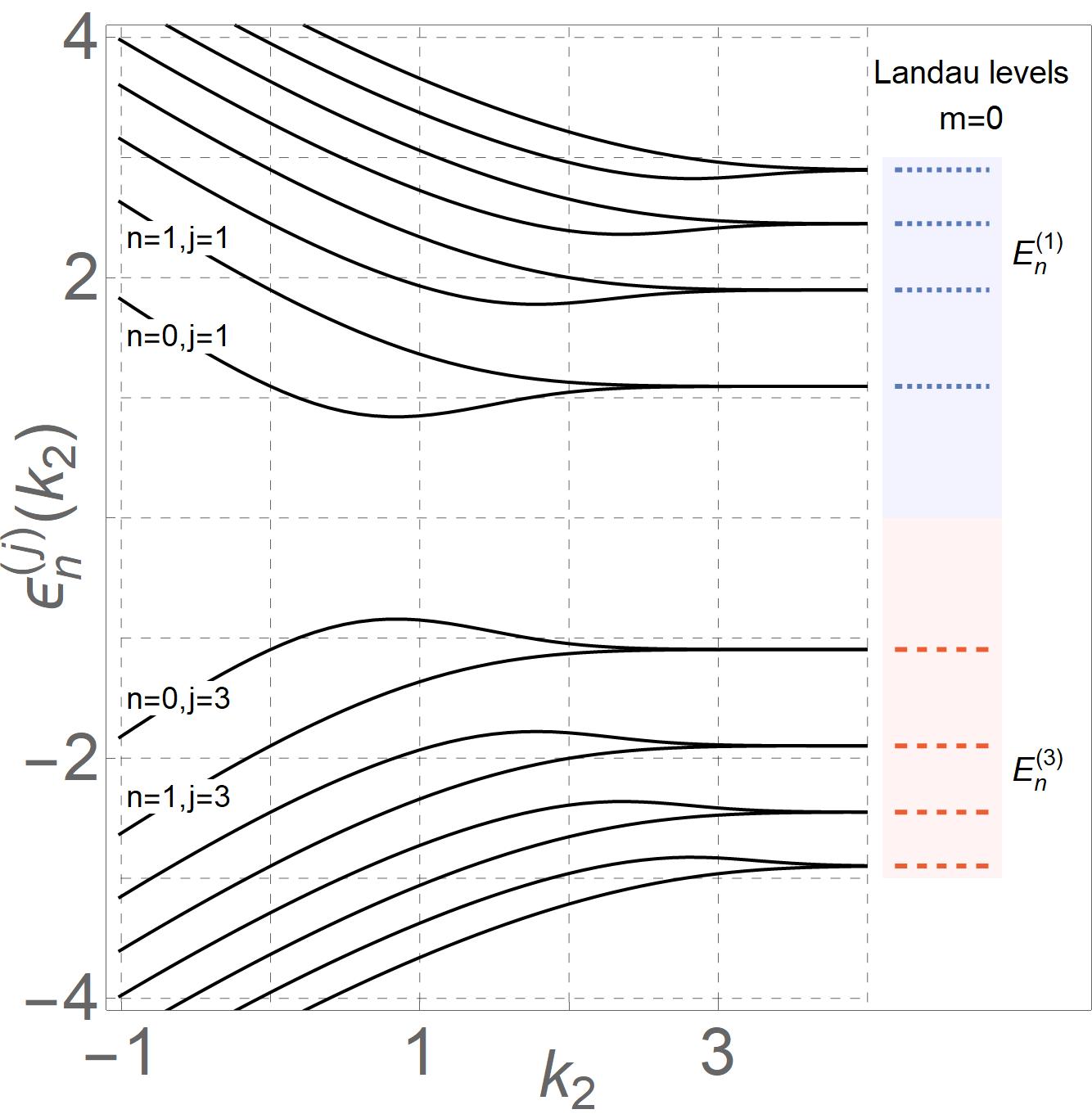}
\label{fig:Eband-1}}
\hspace{3mm}
\subfloat[][$m\neq 0$]{\includegraphics[width=0.6\textwidth]{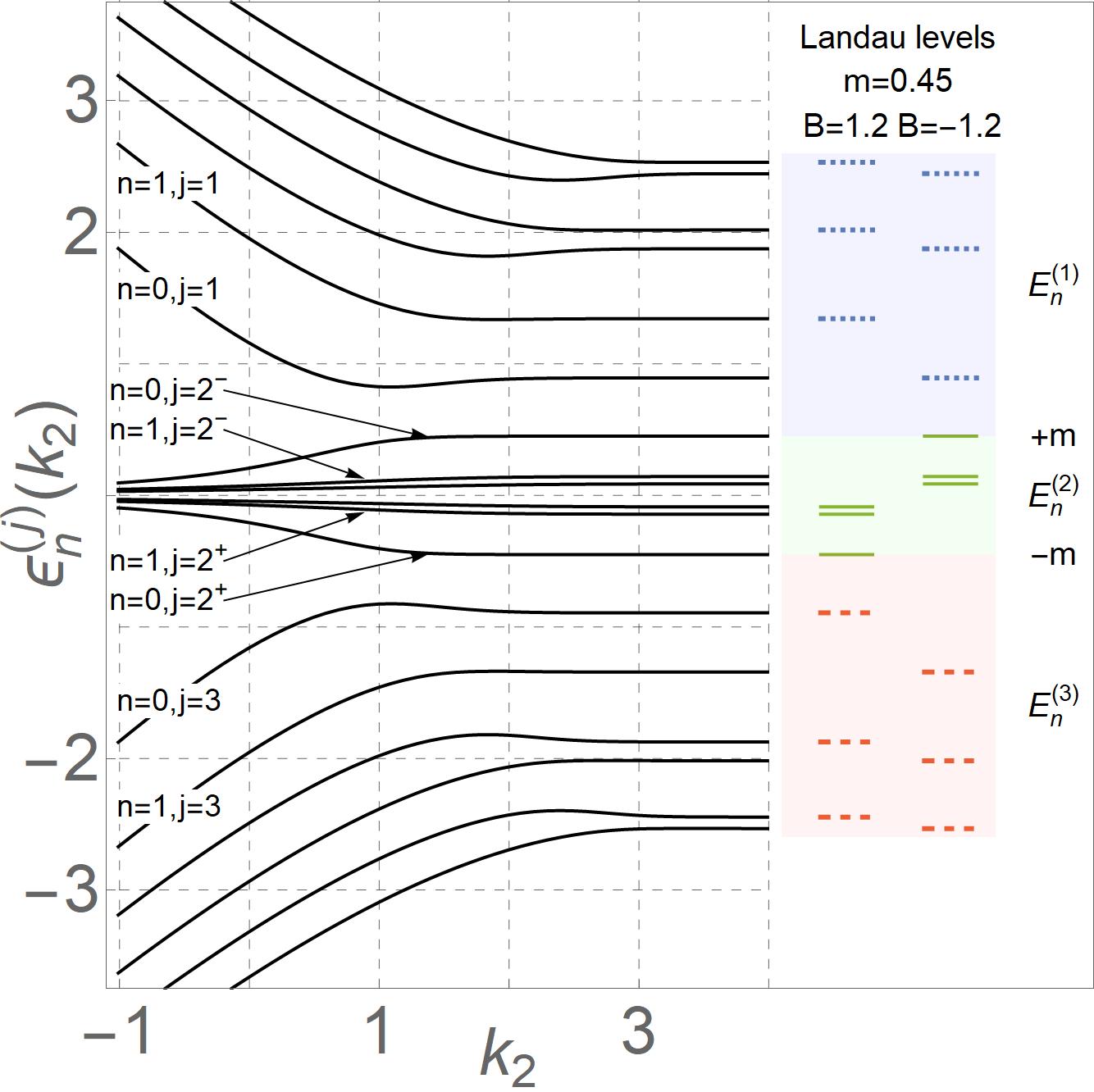}
\label{fig:Eband-2}}
\caption{ Band structure $\epsilon^{(j)}_{n}(k_{2})$ computed from the zeros of~\eqref{secular} for $\hbar=v_{1}=v_{2}=1$, $B=1.2$, where the next-nearest neighbor interaction has been fixed as $m=0$ (a) and $m=0.45$ (b). The levels depicted at the right panel of each figure correspond to the Landau levels $E_{n}^{(j)}$ associated with the homogeneous magnetic fields shown in Figs.~\ref{fig:spec-1}.}
\label{fig:bands}
\end{figure}

In order to verify our latest assertion, we numerically solve the secular equation for $m=0$, and the corresponding behavior of several dispersion bands is depicted in  Fig.~\ref{fig:Eband-1}. We first note that, for large enough values of $k_{2}$, the dispersion bands become constant (dispersionless) and converge to the Landau Levels $E_{n}^{(j)}$ in~\eqref{E-cardano} for $m=0$ which are also depicted in the same figure (see also Fig.~\ref{fig:spec-1}). 

The numerical solutions reveal that two infinite sequences of dispersion bands are generated. In analogy to the index notation used for the Landau levels, we label the set of dispersion bands $\{\epsilon_{n}^{(1)}(k_2)\}_{n=0}^{\infty}$ and $\{\epsilon_{n}^{(3)}(k_2)\}_{n=0}^{\infty}$ as those with with positive ($j=1$) and negative ($j=3$) energies, respectively. Those bands fulfill the symmetry $\epsilon_{n}^{(1)}(k_{2})=-\epsilon_{n}^{(3)}(k_{2})$. Using this notation, we get the asymptotic behavior $\epsilon_{2n}^{(j)}(k_{2}>>1)\sim E_{n}^{(j)}$ and $\epsilon_{2n+1}^{(1)}(k_{2}>>1)\sim E_{n}^{(j)}$ for $j=1,3,$ and the Landau levels $E_{n}^{(j)}$ in~\eqref{E-cardano}.

For $m\neq 0$, dispersion bands are no longer symmetric as the index $\nu_{\pm}$ is not invariant under the reflection $E\rightarrow-E$. Still, for large enough $E$, the index $\nu_{\pm}$ can be approximated as $\nu_{\pm}\vert_{E\rightarrow\infty}\sim\frac{E^{2}}{2\hbar \vert B\vert v_{1}v_{2}}$, which still reveals an asymptotic symmetric behavior. The corresponding numerical solutions of the secular equation depicted in Fig.~\ref{fig:Eband-2} verifies our assertion. Furthermore, in comparison to the case $m=0$, here we obtain two additional infinite dimensional sequences of dispersion bands that emerge inside the free-particle gap $[-m,+m]$. 

\begin{table}
\centering
\subfloat[][$m=0$]{
\begin{tabular}{|clcc|}
\hline
\multicolumn{4}{|c|}{Dispersion band $\epsilon_{n}^{(j)}(k_{2})$}                                                                                                                                                                                                                                                                    \\ \hline
\multicolumn{2}{|c|}{Indexes}                                        & \multicolumn{1}{c|}{\multirow{2}{*}{\begin{tabular}[c]{@{}c@{}}Energy\\ interval\end{tabular}}} & \multirow{2}{*}{\begin{tabular}[c]{@{}c@{}} Behavior \\ for $k_{2}\gg 1$ \end{tabular}} \\ \cline{1-2}
\multicolumn{1}{|c|}{$j$}                  & \multicolumn{1}{l|}{$n$}    & \multicolumn{1}{c|}{}                                                                           &                                                                                                                    \\ \hline
\multicolumn{1}{|c|}{\multirow{2}{*}{1}} & \multicolumn{1}{l|}{$2n$}   & \multicolumn{1}{c|}{\multirow{2}{*}{$(0,+\infty)$}}                                                        & \multirow{2}{*}{$\sim E_{n}^{(1)}$}                                                                                                \\ \cline{2-2}
\multicolumn{1}{|c|}{}                   & \multicolumn{1}{l|}{$2n+1$} & \multicolumn{1}{c|}{}                                                                           &                                                                                                                    \\ \hline
\multicolumn{1}{|c|}{\multirow{2}{*}{3}} & \multicolumn{1}{l|}{$2n$}   & \multicolumn{1}{c|}{\multirow{2}{*}{$(-\infty,0)$}}                                                        & \multirow{2}{*}{$\sim E_{n}^{(3)}$}                                                                                                \\ \cline{2-2}
\multicolumn{1}{|c|}{}                   & \multicolumn{1}{l|}{$2n+1$} & \multicolumn{1}{c|}{}                                                                           &                                                                                                                    \\ \hline
\end{tabular}
}
\subfloat[][$m\neq 0$]{
\begin{tabular}{|clcc|}
\hline
\multicolumn{4}{|c|}{Dispersion band $\epsilon_{n}^{(j)}(k_{2})$}                                                                                                                                                                                                                                                                    \\ \hline
\multicolumn{2}{|c|}{Indexes}                                        & \multicolumn{1}{c|}{\multirow{2}{*}{\begin{tabular}[c]{@{}c@{}}Energy\\ interval\end{tabular}}} & \multirow{2}{*}{\begin{tabular}[c]{@{}c@{}} Behavior \\ for $k_{2}\gg 1$\end{tabular}} \\ \cline{1-2}
\multicolumn{1}{|c|}{$j$}                  & \multicolumn{1}{l|}{$n$}    & \multicolumn{1}{c|}{}                                                                           &                                                                                                                    \\ \hline
\multicolumn{1}{|c|}{\multirow{2}{*}{$1$}} & \multicolumn{1}{l|}{$2n$}   & \multicolumn{1}{c|}{\multirow{2}{*}{$(m,+\infty)$}}                                                        & $\sim E_{n}^{(1)}$ with $B=+\vert B \vert$                                                                                                                \\ \cline{2-2} \cline{4-4} 
\multicolumn{1}{|c|}{}                   & \multicolumn{1}{l|}{$2n+1$} & \multicolumn{1}{c|}{}                                                                           & $\sim E_{n}^{(1)}$ with $B=-\vert B \vert$                                                                                                                                                                                                                               \\ \hline
\multicolumn{1}{|c|}{$2^{+}$} & \multicolumn{1}{l|}{$n$}   & \multicolumn{1}{c|}{$(-m,0)$}                                                        & $\sim E_{n+1}^{(2)}$ with $B=+\vert B\vert$                                                                                                                \\ \hline
\multicolumn{1}{|c|}{$2^{-}$}                   & \multicolumn{1}{l|}{$n$} & \multicolumn{1}{c|}{$(0,m)$}                                                                           & $\sim E_{n+1}^{(2)}$ with $B=-\vert B\vert$                                                                                                                \\ \hline
\multicolumn{1}{|c|}{\multirow{2}{*}{$3$}} & \multicolumn{1}{l|}{$2n$}   & \multicolumn{1}{c|}{\multirow{2}{*}{$(-\infty,-m)$}}                                                        & $\sim E_{n}^{(3)}$ with $B=+\vert B\vert$                                                                                                                \\ \cline{2-2} \cline{4-4} 
\multicolumn{1}{|c|}{}                   & \multicolumn{1}{l|}{$2n+1$} & \multicolumn{1}{c|}{}                                                                           & $\sim E_{n}^{(3)}$ with $B=-\vert B\vert$                                                                                                                \\ \hline
\end{tabular}
}
\caption{Index convention used for the dispersion bands $\epsilon_{n}^{(j)}$ computed from the secular equation~\eqref{secular} and shown in Fig.~\ref{fig:bands} for $m=0$ (a) and $m\neq 0$ (b). The asymptotic values $E_{n}^{(j)}$ correspond to the Landau levels~\eqref{E-cardano} for the homogeneous magnetic field $B=\pm\vert B \vert$.}
\label{tab:snake}
\end{table}

We thus have the four sets of dispersion bands $\{\epsilon_{n}^{(j)}\}_{n=0}^{\infty}$ with $j=1,2^{-},2^{+},3$, which have been arranged such that $\epsilon_{n}^{(1)}\in(+m,\infty)$, $\epsilon_{n}^{(3)}\in(-\infty,-m)$, $\epsilon_{n}^{(2^{+})}\in(-m,0)$, and $\epsilon_{n}^{(2^{-})}\in(0,+m)$. Contrary to the gapless case, here, the bands do not degenerate for large enough values of $k_{2}$, where they become dispersionless and converge to the Landau levels $E_{n}^{(j)}$ associated with both $B=\pm1.2$ and $m\neq 0$ (see Tab.~\ref{tab:snake} for details), as depicted in Fig.~\ref{fig:spec-1}. 

In this form, when the anti-symmetric magnetic field~\eqref{anti-magn} acts on the Lieb lattice, we asymptotically obtain the combined Landau levels $E_{n}^{(j)}$ associated with the homogeneous fields $B=+\vert B\vert$ and $B=-\vert B\vert$, allowing more localized states than either of its the homogeneous symmetric counterparts. Fig.~\ref{fig:Eband-2} also reveals the existence of two energy bands $\epsilon_{0}^{(2^-)}(k_{2})$ and $\epsilon_{0}^{(2^+)}(k_{2})$ that asymptotically approach to $m$ and $-m$, respectively. For the homogeneous magnetic field case discussed in Sec.~\ref{sec:landau}, the eigenvalues $E=\pm m$ are discarded from the discrete spectrum due to the non-square-integrability of the corresponding eigensolutions. Nevertheless, the energy bands never reach the forbidden values\footnote{Recall that such values are forbidden as the components $\psi_{B}$ and $\psi_{C}$ in~\eqref{psi-q} diverge.} $\pm m$ for the anti-symmetric magnetic field setup, and thus the corresponding eigensolutions are always square-integrable. Thus, those two energy bands generate localized states for arbitrary values of $k_{2}$, including $k_{2}\gg 1$.

\begin{figure}
\centering
\subfloat[][$m=0$]{\includegraphics[width=0.4\textwidth]{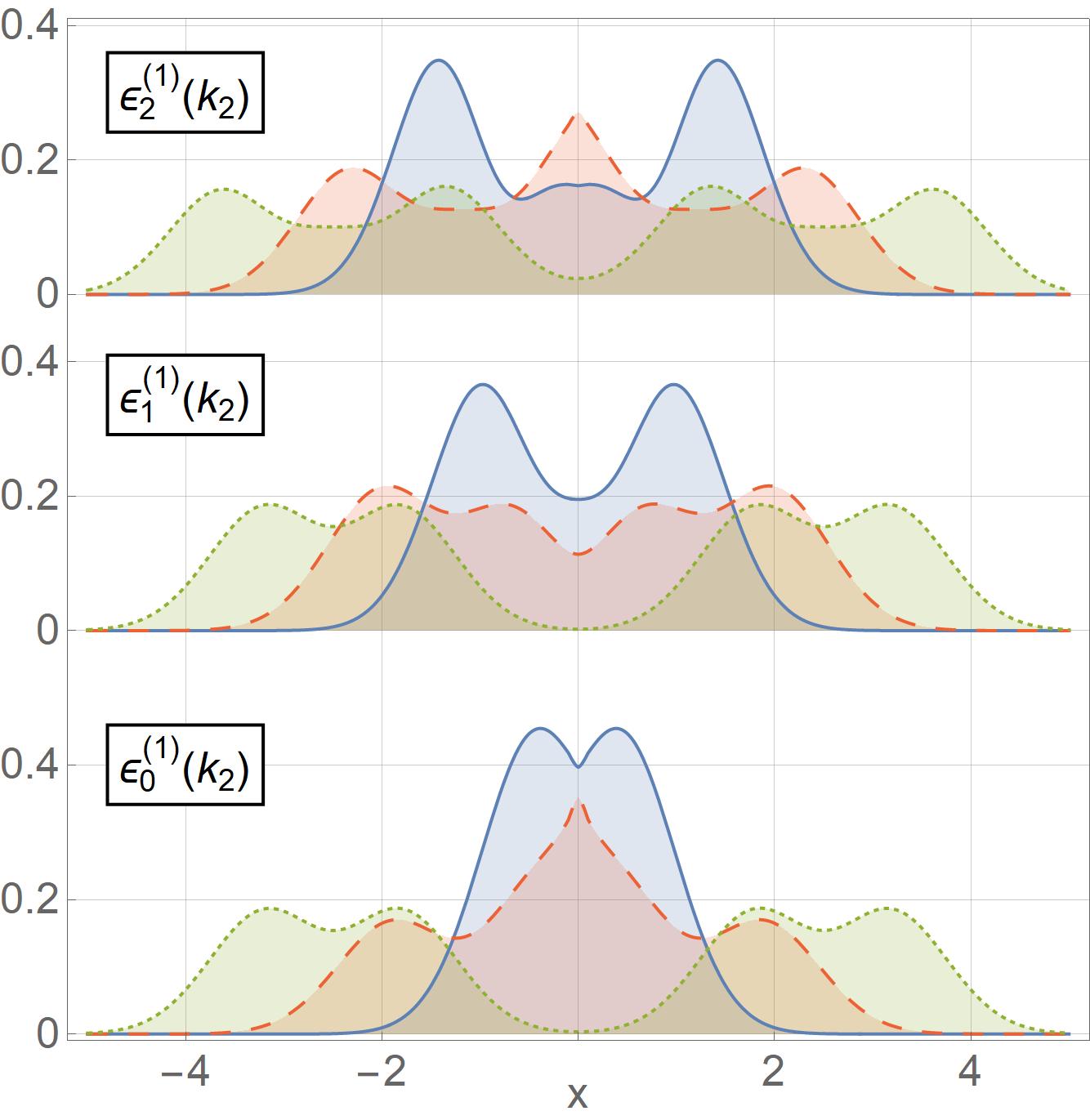}
\label{fig:density-1}}
\hspace{3mm}
\subfloat[][$m=0.45$]{\includegraphics[width=0.4\textwidth]{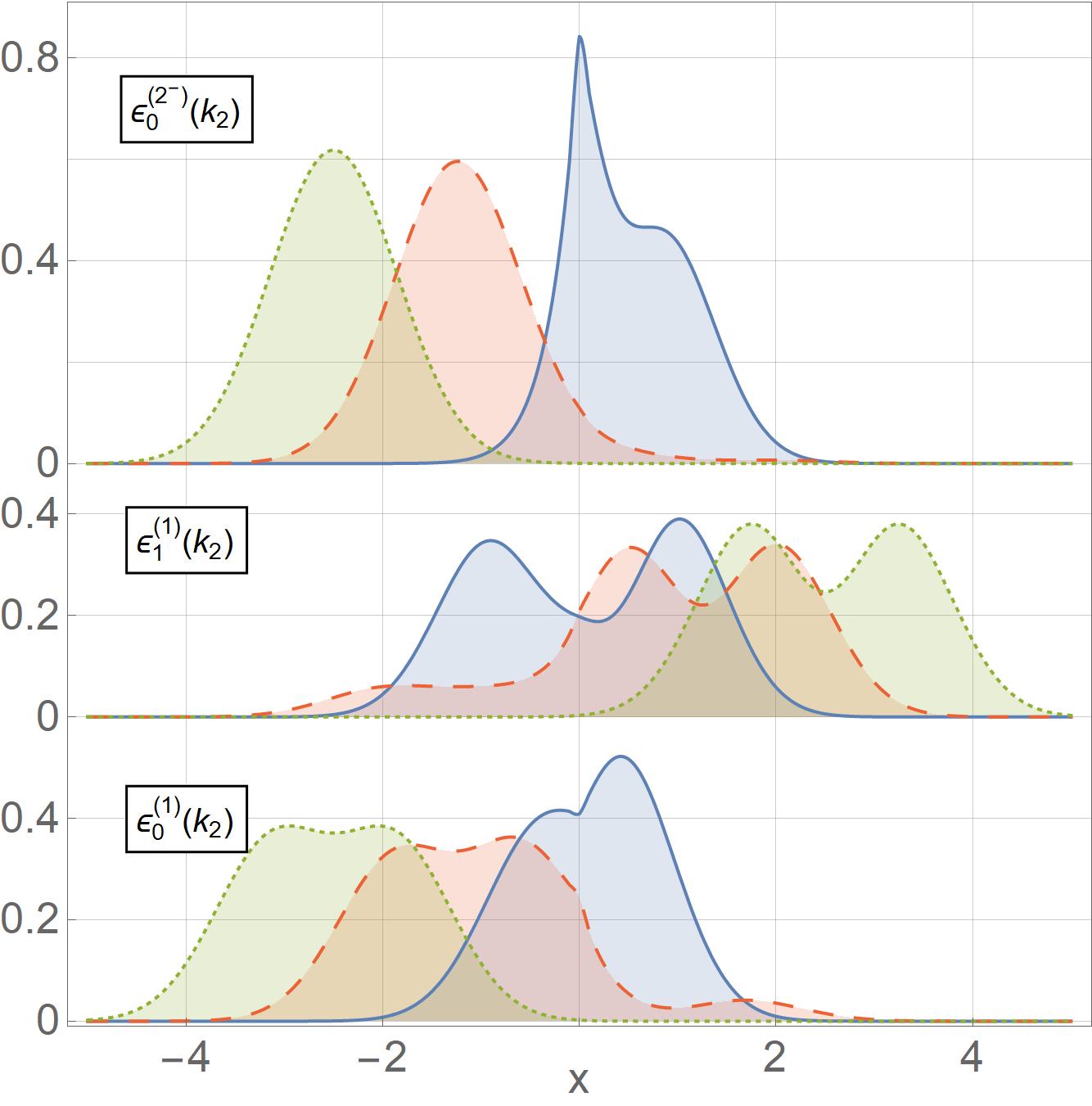}
\label{fig:density-2}}
\caption{Probability density $\rho=\boldsymbol{\Psi}^{\dagger}\boldsymbol{\Psi}$ for the localized eigensolutions associated with the bands $\epsilon^{(j)}_{n}(k_{2})$. The parameters have been fixed as in Fig.~\ref{fig:bands} with $k_{2}=-0.5$ (blue-solid), $k_{2}=1.5$ (red-dashed), and $k_{2}=3$ (green-dotted), together with $m=0$ (a) and $m=0.45$ (b).}
\label{fig:density}
\end{figure}

In Fig.~\ref{fig:density} we depict the corresponding probability density $\rho=\boldsymbol{\Psi}^{\dagger}\boldsymbol{\Psi}$ associated with the $\epsilon_{n}^{(j)}(k_{2})$ for several values of $k_{2}$ and $n,j$. Particularly, for the gapless case $m=0$, we have a symmetric distribution with respect to the discontinuity in the magnetic field, $x=0$, where the localized states associated with $\epsilon_{n}^{(1)}(k_{2})$ are highly localized around $x=0$ for $k_{2}=-0.5$. The same behavior is obtained for $\epsilon_{n}^{(3)}(k_{2})=-\epsilon_{n}^{(1)}(k_{2})$. For larger values of $k_{2}$, the distribution equally spreads across the regions of positive ($x>0$) and negative ($x<0$) magnetic field amplitude so that, for higher $n$, electrons are most likely to be localized away from $x=0$. Note that $\rho$ for $\epsilon_{0}^{(1)}(k_{2})$ and $\epsilon_{1}^{(1)}(k_{2})$ converge to the same distribution for $k_{2}\gg 1$, as expected as their corresponding bands degenerate.

If the band-gap opens ($m\neq 0$), the probability distribution $\rho$ loses its symmetric behavior, which induces a bias in the probability distribution across the positive or negative magnetic field regions, allowing privileged zones where electrons are primarily localized. This can be seen in Fig.~\ref{fig:density-2}, where electrons are likely to be found in the positive magnetic field region for $k_{2}=-0.5$. In contradistinction to the gapless case, here, one may notice that $\rho$ associated with $\epsilon_{0}^{(1)}(k_{2})$ and $\epsilon_{1}^{(1)}(k_{2})$ deviate from each other. On the other hand, for $k_{2}=1.5$ and $k_{2}=3$, electrons predominantly pile on the negative magnetic field region for the bands $\epsilon_{0}^{(1)}(k_{2})$ and $\epsilon_{0}^{(2^-)}(k_{2})$, whereas electrons localize on the positive magnetic field region for $\epsilon_{1}^{(1)}(k_{2})$. 

\begin{figure}
\centering
\subfloat[][$m=0$, $\epsilon_{0}^{(1)}(k_{2})$]{\includegraphics[width=0.48\textwidth]{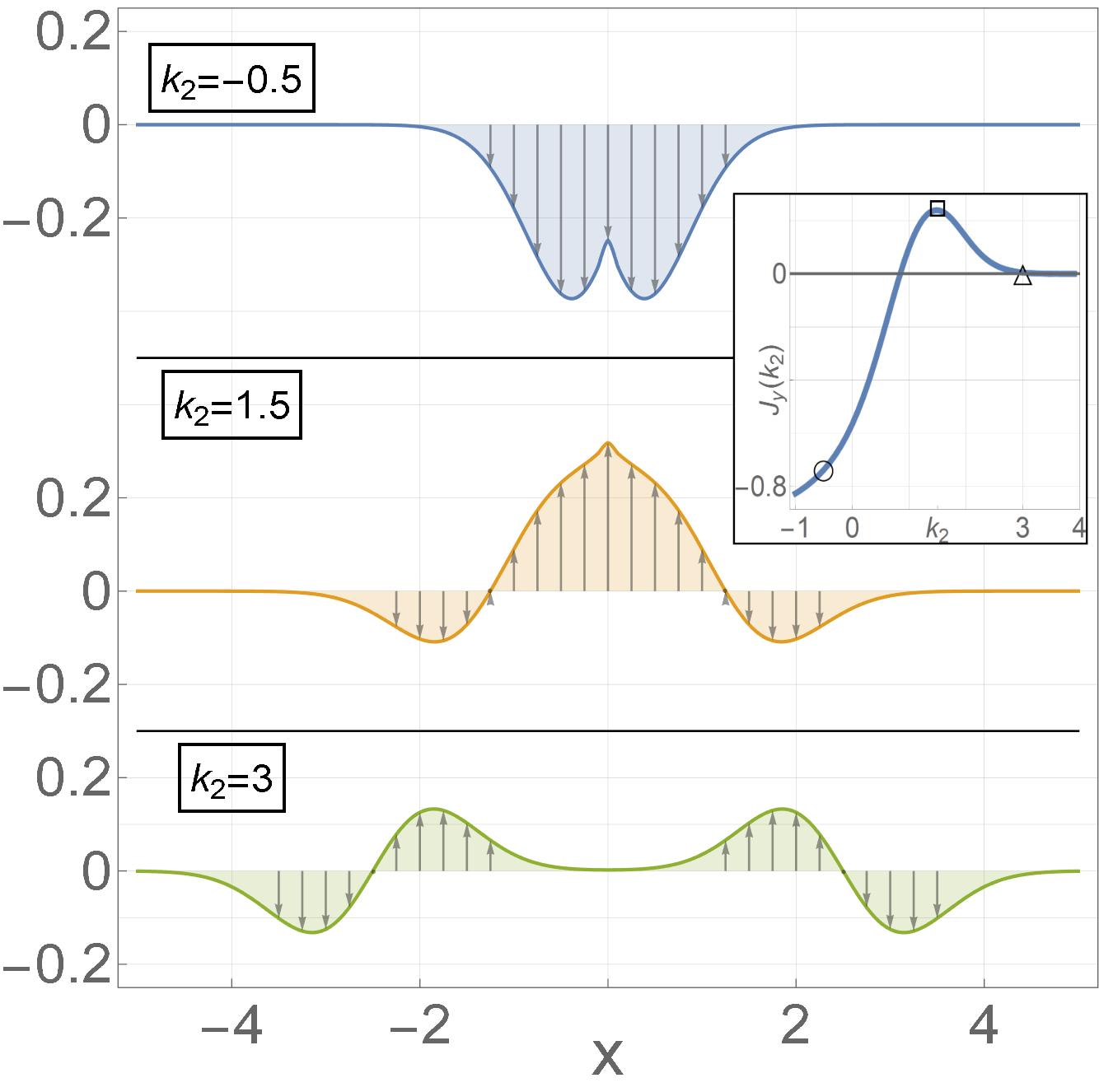}
\label{fig:current-1}}
\hspace{2mm}
\subfloat[][$m=0.45$, $\epsilon_{0}^{(2^-)}(k_{2})$]{\includegraphics[width=0.48\textwidth]{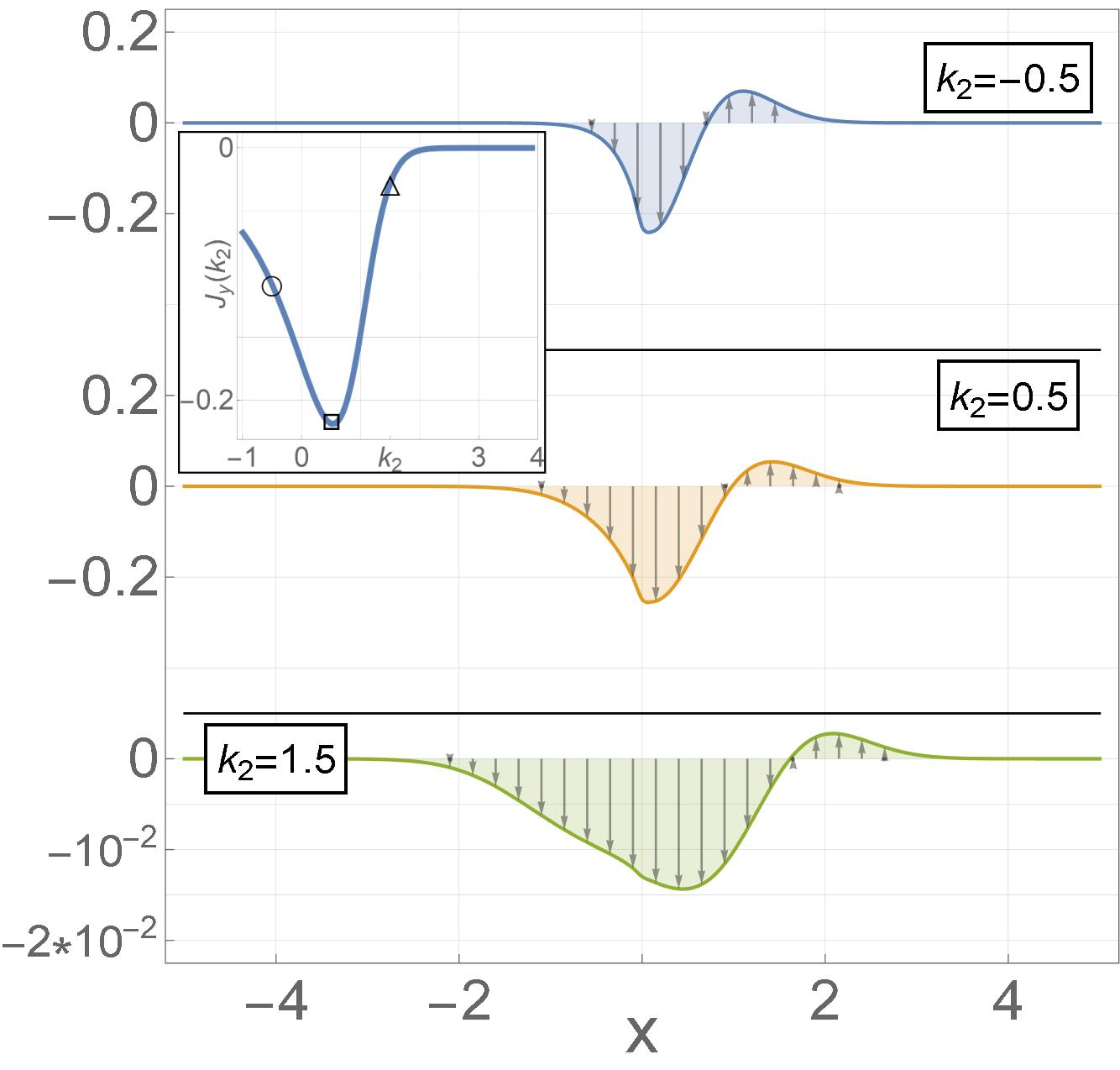}
\label{fig:current-2}}
\caption{Current density in the $\hat{y}$ direction, $j_{y}(x,k_{2})$, as a function of the position for the localized eigensolutions associated with the bands $\epsilon^{(j)}_{n}(k_{2})$. We have considered $m=0$ and $\epsilon_{0}^{(1)}(k_{2})$ (a), and $m=0.45$ and $\epsilon_{0}^{(2)}(k_{2})$ (b). $k_{2}$ has been specified in each panel and the rest of parameters are fixed as in Fig.~\ref{fig:bands}.  The insets depict the associated group velocity $v_{y}(k_{2})$, or equivalently the net current $J_{2}(k_{2})$, computed from~\eqref{vel-group}. The circle, square, and triangle marks denote the respective values of $k_{2}$ in the sets $\{k_{2}=-0.5,k_{2}=1.5,k_{2}=3\}$ (a) and $\{k_{2}=-0.5,k_{2}=0.5,k_{2}=1.5\}$.}
\label{fig:current}
\end{figure}

The localized states with eigenvalues $\epsilon_{n}^{(j)}(k_{2})$ carry an intrinsic probability current, in accordance with the classical picture where the snake states propagate along the interface. Similarly to the homogeneous magnetic field case, the current distribution in the $\hat{x}$ direction is exactly zero. Nevertheless, in the anti-symmetric case, the currents in the $\hat{y}$ direction do not longer compensate each other and a net effect can be observed. This can verified through the group velocity in~\eqref{vel-group}, which is equivalent to the net current $J_{y}(k_2)=\int_{\mathbb{R}}dx\, j_{y}(x;k_{2})=\partial_{k_{2}}\epsilon_{n}^{(j)}(k_{2})$ and can be extracted directly from the bands depicted in Fig.~\eqref{fig:bands}. 

Therefore, each of the bands $\epsilon_{n}^{(j)}(k_{2})$ carries current for arbitrary values of $k_{2}$, except for $k_{2}\gg 1$, where the net current contribution is approximately null. This is due to the fact that the bands are asymptotically constant and approach the Landau levels $E_{n}^{(j)}$. To verify the latter, we depict the probability current associated with the bands $\epsilon_{n}^{(j)}$ in Fig.~\ref{fig:current}. Particularly, Fig.~\ref{fig:current-1} shows density currents for the gapless case and the lower positive band $\epsilon_{0}^{(1)}$. There, we may note that indeed, for negative and small enough positive values of $k_{2}$, the density currents are unbalanced, leading to a net negative and positive current for $k_{2}=-0.5$ and $k_{2}=1.5$, respectively. For $k_{2}=3$, the band approaches to the corresponding Landau level, and the currents become approximately balanced, vanishing any net effect on the current. In Fig.~\ref{fig:current-2}, we show instead the behavior of the current for the gapped case, $m=0.45$, and the band appearing inside the gap $\epsilon_{0}^{(2^-)}$. We have a similar case, although the net currents are always negative.

%\cite{Oro08,Gho08}

\section{Remarks about other phase-hopping $\lambda$}
\label{sec:lambda}
%\subsection{Vicinity of $\lambda=\frac{\pi}{2}$}
$\bullet$ We analyzed the system for $\lambda=\frac{\pi}{2}$ in the vicinity of $\vec{K}_{\operatorname{I}}$ where the energy $w_+$ has its minimum. When $\lambda\neq 0$, analytical treatment of minima of (\ref{genericE}) for generic values of $\lambda$ is not feasible. Nevertheless, it is reasonable to expect that for $\lambda=\frac{\pi}{2}+\delta$ with $\delta <<1$, the minimum of $w_+$ stays in the vicinity of $\vec{K}_{\operatorname{I}}$ or it even does not move at all. In either case, when we expand the tight-binding Hamiltonian (\ref{TB-2}) at $\vec{K}_{\operatorname{I}}$, we get 
\begin{equation}
\mathcal{H}_{\operatorname{I}}(\lambda,\vec{k}) =
\begin{pmatrix}
0 &2a t_1 k_x & 2a t_2 k_y \\
2a t_1 k_x & 0 & 4it_3\sin\lambda\\
2a t_2 k_y & -4it_3\sin\lambda & 0
\end{pmatrix}
.
\label{TB-2DP1}
\end{equation}
In comparison to $\mathcal{H}_{\operatorname{I}}(\vec{k})$, the coupling constant $t_3$ is replaced by $t_3\rightarrow t_3\sin\lambda$,
\begin{equation}
\mathcal{H}_{\operatorname{I}}(\lambda,\vec{k})=\mathcal{H}_{\operatorname{I}}(\vec{k})\vert_{t_3\rightarrow t_3\sin\lambda}
\end{equation}
It suggests that the change of $\lambda$ weakens the influence of the NNN interaction $t_3$ on dynamics of the quasi-particles.

%\subsection{Upper and lower flat bands}
$\bullet$ It is worth to remark that additional values of $\lambda$ exist so that the dispersion relations support one flat band in the free-particle case ($\vec{A}=0$). Indeed, if we fix $\lambda=\frac{\pi}{4}$ and $\lambda=\frac{3\pi}{4}$, together with $t_{3}=t$ and $t_{1}=t_{2}=\sqrt{2}t$, we simplify the characteristic equation~\eqref{band-pi2} so that we get the solutions
\begin{align}
&w_{\frac{\pi}{4},f}(\vec{k})=2\sqrt{2}t , \quad &w_{\frac{\pi}{4},\pm}(\vec{k})=\sqrt{2}t\left(-1\pm\sqrt{1+8\cos^{2}(ak_{x})\cos^{2}(ak_{y})}\right) , 
\label{pi4}\\
&w_{\frac{3\pi}{4},f}(\vec{k})=-2\sqrt{2}t , \quad &w_{\frac{3\pi}{4},\pm}(\vec{k})=\sqrt{2}t\left(1\pm\sqrt{1+8\cos^{2}(ak_{x})\cos^{2}(ak_{y})}\right) .
\label{3pi4} 
\end{align}
Clearly, this case is more restrictive to that discussed in Sec.~\ref{sec:spin1} as we impose a constraint to all the hopping parameters.

We thus generate an upper (lower) flat band for $\lambda=\frac{\pi}{4}$ ($\lambda=\frac{3\pi}{4}$) and two dispersion bands below (above) it. Further analysis shows that, for $\lambda=\frac{\pi}{4}$ (or $\lambda=\frac{3\pi}{4}$), the upper (lower) flat band intercepts the $w_{\frac{\pi}{4},+}(\vec{k})$ (or $w_{\frac{3\pi}{4},-}(\vec{k})$) at the origin of the Brilloin zone $\vec{K}_{0}=(0,0)$. 

\begin{figure}
\centering
\subfloat[][$\lambda=\frac{\pi}{4}$]{\includegraphics[width=0.35\textwidth]{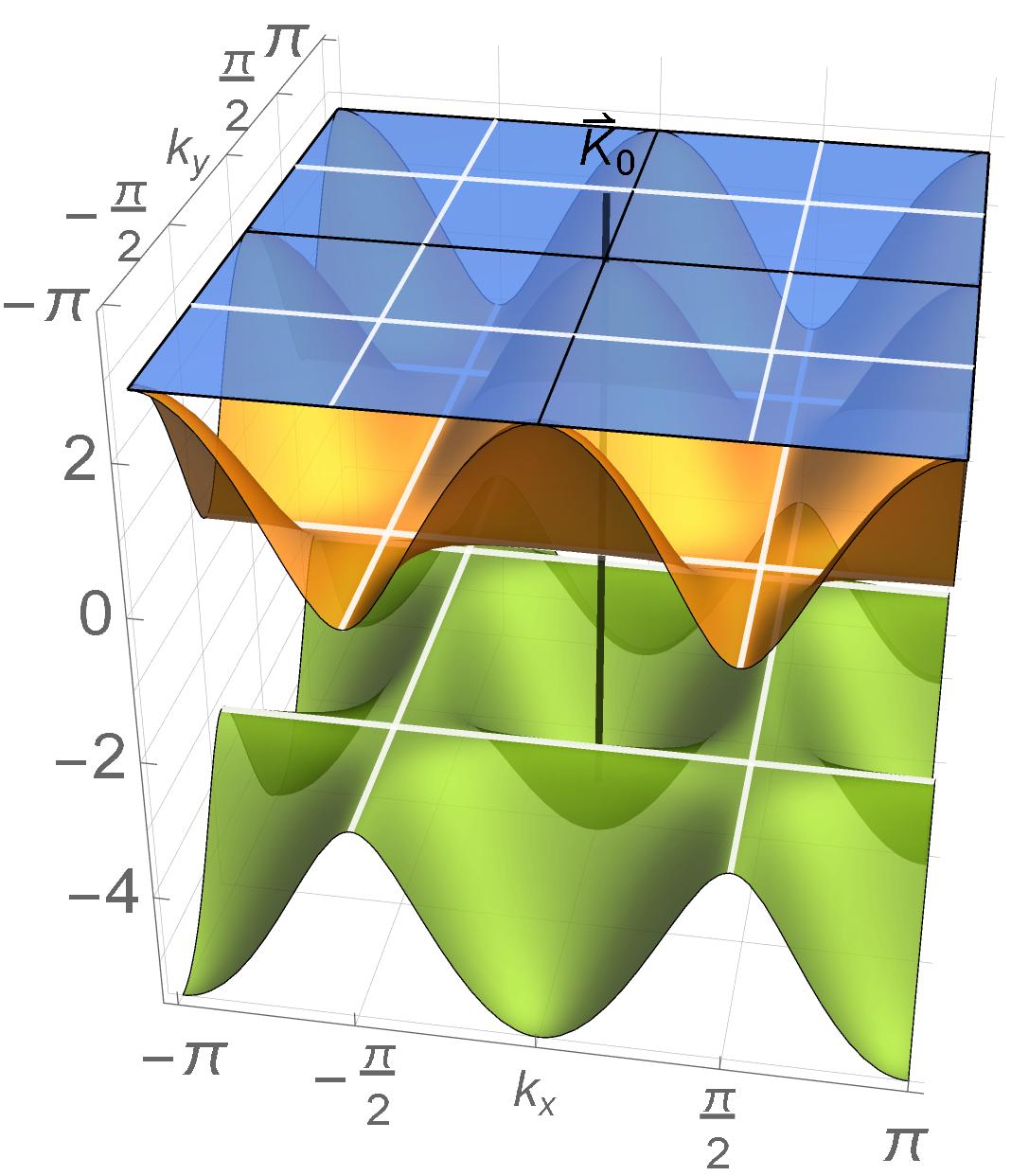}
\label{fig:flat1}}
\hspace{2mm}
\subfloat[][$\lambda=\frac{3\pi}{4}$]{\includegraphics[width=0.35\textwidth]{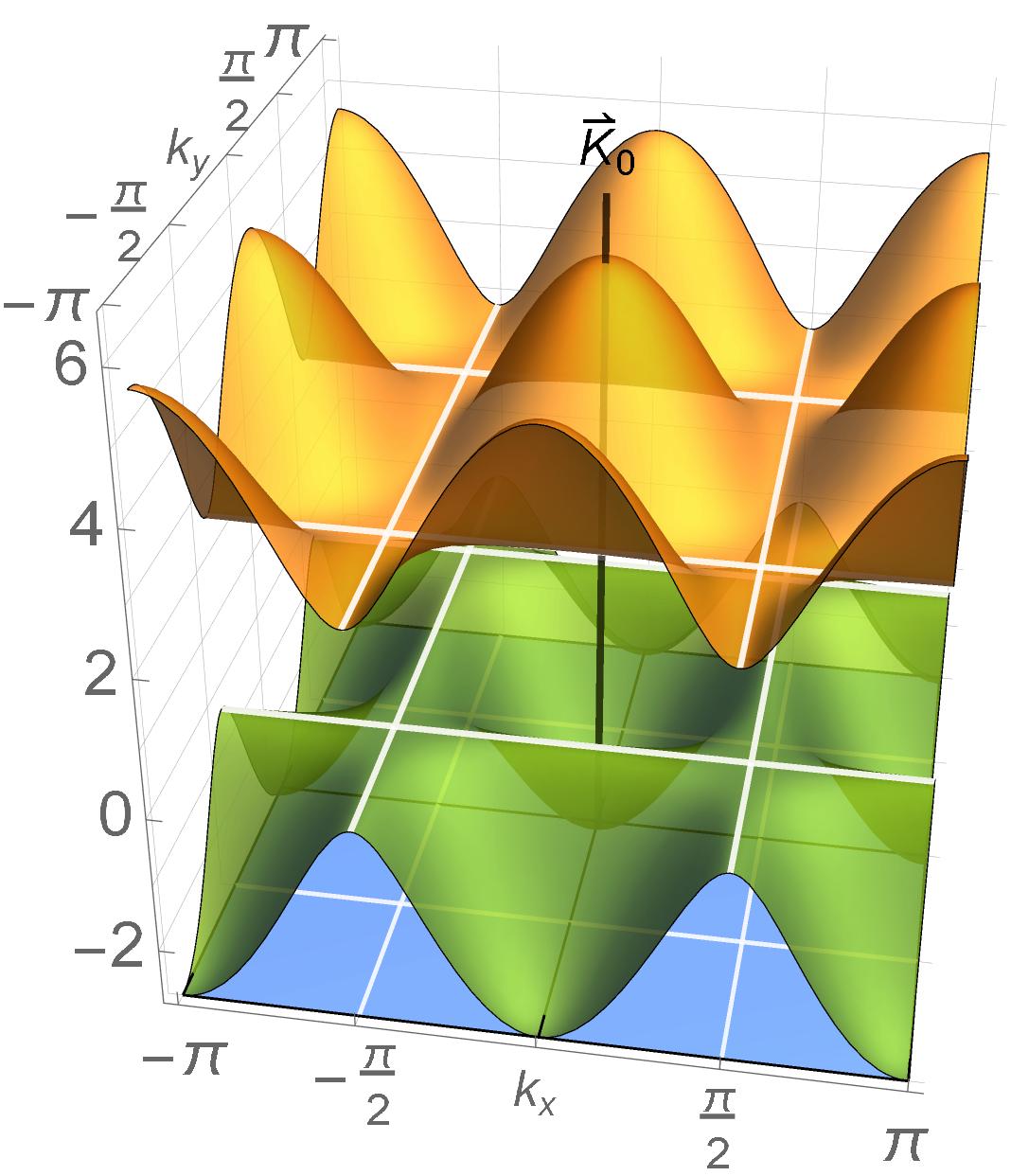}
\label{fig:flat2}}
\caption{Flat and dispersion bands in~\eqref{pi4}-\eqref{3pi4} for $t=1$. The white lines denote the set of points generated by $\vec{K}_{f_{1}}$ and $\vec{K}_{f_{2}}$, and $\vec{K}_{0}$ marks the origin in the $k$-space.}
\label{fig:flat}
\end{figure}

On the other hand, the dispersion bands have a gap of $2\sqrt{2}t$ for both values of $\lambda$, which occurs at the values of $\vec{k}$ that minimizes $w_{\frac{\pi}{4},+}(\vec{k})$ and maximizes $w_{\frac{\pi}{4},-}(\vec{k})$ simultaneously. Recall that, for $\lambda=\frac{\pi}{2}$, those points form the discrete set labeled in Tab.~\ref{tab:DiracP} plus some invariant translations in the $\vec{k}$-space. In contradistinction, for both $\lambda=\frac{\pi}{4}$ and $\lambda=\frac{3\pi}{4}$, the gap emerges for a continuous (non-countable) set of points in the Brillouin zone, constructed through the unequivalent vectors 
\begin{equation}
\vec{K}_{f_1}=\left( \frac{\pi}{2}, \kappa_{2} \right), \quad \vec{K}_{f_2}=\left( \kappa_{1},\frac{\pi}{2} \right), \quad \kappa_{1,2}\in\left(-\frac{\pi}{2},\frac{\pi}{2}\right) .
\label{pi4-plane}
\end{equation}
This generates two unequivalent lines of points in the $\vec{k}$-space (plus some invariant translations), which are depicted in Fig.~\ref{fig:flat1} as white lines marked on the dispersion bands. Analogous results are recovered for $\lambda=\frac{3\pi}{4}$ in Fig.~\ref{fig:flat2}.

Although the tight-binding Hamiltonian $\mathcal{H}(\vec{k})$ in~\eqref{TB-2} can be expanded around the intercepting point $\vec{K}_{0}$ on the flat band, further analysis is required for the continuous set of points $\vec{K}_{f_{1}}$ and $\vec{K}_{f_{2}}$ that generate the gap. This will be discussed elsewhere.

%---------------------------------------> Section
\section{Concluding remarks}
This work has studied the three-band structure of a Lieb lattice with distinct hopping amplitudes for near-neighbors and a complex phase-hopping for next-nearest neighbors interactions. For a pure imaginary NNN interaction ($\lambda=\pi/2$), one can identify three different Dirac valleys in which the effective tight-binding Hamiltonian reduces to an effective pseudo-spin-1 Dirac equation. The position of such valleys depends strongly on a set of relations among the hopping amplitudes $t_{1,2,3}$. One thus identifies four different cases, which are classified in Tab.~\ref{tab:DiracP}. For $t_{1}=t_{2}$, the second and third cases in Tab.~\ref{tab:DiracP} disappear, and one thus has only two different configurations of Dirac valleys. This shows the versatility achieved by allowing the hopping amplitudes to be all different.

In the presence of external homogeneous magnetic fields, the Dirac equation can be solved exactly, and the corresponding Landau levels $E_{n}^{(j)}$ can be computed exactly in general for arbitrary hopping parameter amplitudes $t_{1,2,3}$ and specific phase hopping $\lambda=\pi/2$. In contradistinction to the Landau in graphene, here we have three infinite-dimensional sequences of eigenvalues, two of which are reminiscent of the graphene case, and a new sequence that is strictly associated with the intermediate flat band. The latter sequence degenerates into a single unphysical level when the free-particle band gap closes ($t_{3}=0$), and the remaining levels are akin to Landau levels in graphene.

We obtain current-carrying localized states, the snake states, for anti-symmetric magnetic fields. For $t_{3}=0$, the dispersion relations $\epsilon_{n}^{(j)}(k_{2})$ asymptotically degenerate by pairs and converge to the corresponding Landau levels $E_{n}^{(j)}$ of the homogeneous field case. Remarkably, for $t_{3}\neq 0$, the dispersion relations do not degenerate asymptotically and converge to the Landau levels associated with both $B=+\vert B\vert$ and $B=-\vert B\vert$. In this regard, we have a lattice with stationary levels whose spectral information may be thought of as $\sigma(H_{\operatorname{I}}(k_{2}))\vert_{\\k_{2}\gg 1}\sim\sigma(H_{\operatorname{I}})\vert_{B=+\vert B \vert}\cup \sigma(H_{\operatorname{I}})\vert_{B=-\vert B \vert}\cup \{-m+\delta, m-\delta\}$ with $\sigma(H_{\operatorname{I}})$ given in~\eqref{Landau-Lieb} and $\delta\ll 1$. Remark that the energies $E=\pm m$ cannot be associated with finite-norm eigensolutions. Still, here, the dispersion relations converge to such values only asymptotically, and thus those energies are never physically achieved. For that reason, we have included values in the vicinity of $\pm m$, obtained for large enough $k_{2}$.

Interestingly, for the free-particle case, the flat band can be moved and placed either on top or at the bottom of the other two dispersion bands by tuning the phase hopping $\lambda$. However, further strict constraints are required among the hopping amplitudes to achieve this band structure. In such a case, the flat band always intercepts one of the dispersion bands, and the dispersion bands become gapped. In this case, the set of points defining the Dirac valleys form a continuum in the $k$-space, as seen in Fig.~\ref{fig:flat}. Thus the proper classification of Dirac points and the expansion of the corresponding Hamiltonian requires further analysis, which deserves attention by itself and will be discussed in full detail elsewhere.

%---------------------------------------> Section

%---------------------------------------> Section
\section*{Acknowledgments}
K.Z. acknowledges the support from the project ``Physicists on the move II'' (KINE\'O II), Czech Republic, Grant No. CZ.02.2.69/0.0/0.0/18 053/0017163. 
%and the support of Consejo Nacional de Ciencia y Tecnolog\'ia, Mexico, grant number A1-S-24569.

\appendix
\setcounter{section}{0}
\section{Cardano formulas}
\label{sec:cardano}
\renewcommand{\thesection}{A-\arabic{section}}
% redefine the command that creates the equation no.
%\setcounter{section}{0}  % reset counter 
\renewcommand{\theequation}{A-\arabic{equation}}
% redefine the command that creates the equation no.
\setcounter{equation}{0}  % reset counter

Be the general third-order polynomial equation
\begin{equation}
z^{3}+a_{2}z^{2}+a_{1}z+a_{0}=0 ,
\label{p3c}
\end{equation}
with $a_{j}\in\mathbb{R}$ for $j=0,1,2$. It is straightforward to realize that the reparametrization
\begin{equation}
z=x-\frac{a_{2}}{3} , \quad p=a_{1}-\frac{a_{2}^{2}}{3} , \quad q=a_{0}-\frac{a_{1}a_{2}}{3}+\frac{2}{27}a_{2}^{3} , \quad p,q\in\mathbb{R} .
\end{equation}
reduces~\eqref{p3c} into the incomplete cubic equation
\begin{equation}
x^{3}+px+q=0 .
\label{p3}
\end{equation}
%In this form, the solutions of~\eqref{p3} can be transformed into solutions of~\eqref{p3c}, and vice versa. 
The most general solution for~\eqref{p3} is given by
\begin{equation}\label{genericEe}
x=\left(-\frac{q}{2}+\sqrt{\frac{q^2}{4}+\frac{p^3}{27}}\right)^{\frac{1}{3}}+\left(-\frac{q}{2}-\sqrt{\frac{q^2}{4}+\frac{p^3}{27}}\right)^{\frac{1}{3}}
\end{equation}
provided that the cubic roots are selected such that  
\begin{equation}
\left(-\frac{q}{2}+\sqrt{\frac{q^2}{4}+\frac{p^3}{27}}\right)^{\frac{1}{3}}\left(-\frac{q}{2}-\sqrt{\frac{q^2}{4}+\frac{p^3}{27}}\right)^{\frac{1}{3}}=-\frac{p}{3}.
\end{equation}

In order to determine whether the solutions of~\eqref{p3} are simple or degenerate, as well as real or complex, one has to analyze the discriminant
\begin{equation}
\Delta:=4p^{3}+27q^{2} .
\end{equation}
In this form, if $\Delta\leq 0$, we can ensure the existence of three real solutions. Particularly, for $\Delta<0$, all solutions are different, whereas for $\Delta=0$, at least two solutions are equal (degenerate). Regardless the case, we have the three solutions as~\cite{Olv10}
\begin{equation}
x_{j}=A\sin \left( a + (j-1)\frac{2\pi}{3}\right) , \quad A=\sqrt{-\frac{4p}{3}} , \quad \sin(3a)=\frac{4q}{A^{3}} , \quad j=1,2,3.
\label{cubic}
\end{equation}
This can be conveniently rewritten, by using elementary trigonometric relations, as
\begin{equation}
x_{1}=A\cos\left( \frac{\theta+(j-1)2\pi}{3} \right) , \quad \theta=\arccos\left( -\frac{4q}{A^{3}} \right) , \quad j=1,2,3.
\label{cubic-f}
\end{equation}

Although the latter applies to the eigenvalues computed for the homogeneous field in Sec.~\ref{sec:landau}, it is worth to mention that there may be one complex or two mutually complex-conjugated solutions for $\Delta>0$. This case will not be discussed any further as it is not required for our calculations. See~\cite{Olv10} for further details.

%---------------------------------------> Bibliography

\end{document}